\documentclass[12pt]{iopart}
\usepackage{cite}
\usepackage[top=3cm, bottom=3cm, left=3cm, right=3cm]{geometry} 
\usepackage[T1]{fontenc}
\usepackage{graphicx}
\expandafter\let\csname equation*\endcsname\relax
\expandafter\let\csname endequation*\endcsname\relax
\usepackage{amsmath}
\usepackage{amsthm}
\usepackage{amsfonts}
\usepackage{amssymb}
\usepackage{epstopdf}

\usepackage{color}
\newcommand{\defeq}{\mathrel{\mathop:}=}
\newcommand{\Ref}[1]{(\ref{#1})}
\newcommand{\bg}{\begin}
\newcommand{\bd}[1]{\ensuremath{\boldsymbol{#1}}}
\newcommand{\lt}{\left }
\newcommand{\rt}{\right}
\newcommand{\bgmx}{\begin{bmatrix}}
\newcommand{\emx}{\end{bmatrix}}
\newcommand{\dd}{\ensuremath{\mathrm{d}}}
\newcommand{\f}[2]{\ensuremath{\frac{#1}{#2}}}
\newcommand{\deq}{\ensuremath{\overset{d}{=}}}
\newcommand{\I}{\ensuremath{\mathrm{i}}}
\newcommand{\R}{\ensuremath{\mathbb{R}}}
\newcommand{\N}{\ensuremath{\mathbb{N}}}
\newcommand{\pr}{\mathrm{P}}
\newcommand{\E}{\mathbb{E}}

\newcommand{\bgeq}{\begin{equation}}
\newcommand{\eeq}{\end{equation}}

\begin{document}

\title[Superstatistical generalised Langevin equation]{Superstatistical
generalised Langevin equation: non-Gaussian viscoelastic anomalous diffusion}

\author{Jakub \'Sl\k{e}zak$^{\dagger,\flat}$, Ralf Metzler$^\flat$, Marcin
Magdziarz$^\dagger$}
\address{$\dagger$ Hugo Steinhaus Center, Faculty of Pure and Applied Mathematics, Wroc{\l}aw University of Science and Technology, Wybrze{\.z}e Wyspia{\'n}skiego 27, 50-370
Wroc{\l}aw, Poland\\
$\flat$ Institute for Physics \& Astronomy, University of Potsdam,
Karl-Liebknecht-Stra{\ss}e 24/25, 14476 Potsdam-Golm, Germany}

\begin{abstract}
Recent advances in single particle tracking and supercomputing techniques
demonstrate the emergence of normal or anomalous, viscoelastic diffusion
in conjunction with non-Gaussian distributions in soft, biological, and
active matter systems. We here formulate a stochastic model based on a
generalised Langevin equation in which non-Gaussian shapes of the probability density function and normal or anomalous diffusion have a common origin, namely a random parametrisation of the stochastic force. We perform a
detailed analytical analysis demonstrating how various types of parameter
distributions for the memory kernel result in the exponential, power law,
or power-log law tails of the memory functions.
The studied system is also shown to exhibit a further unusual property: the
velocity has a Gaussian one point probability density but non-Gaussian joint
distributions. This behaviour is reflected in relaxation from Gaussian to
non-Gaussian distribution observed for the position variable. We show that
our theoretical results are in excellent agreement with Monte Carlo simulations.
\end{abstract}

\section{Introduction}

At the beginning of 20th century the works of Einstein, Smoluchowski, Langevin
and Wiener \cite{einstein,smol,langevinEng,wiener} opened a new chapter of
quantitative understanding of physics, chemistry, and mathematics by laying
down the foundations for what we now call the theory of stochastic processes.
Their goal was to provide descriptions of various aspects of diffusive motion,
which were observed even in ancient times, for instance, by Roman poet Lucretius
\cite{lucretius}. However, it was the groundbreaking experiments of Brown in the
19th century that brought this topic into scientific attention.

Two fundamental properties are commonly encountered in observed diffusive motion:
(i) The mean squared displacement (MSD) of the particle position $X$ grows linearly 
with time,
\begin{equation}
\label{eq:msd}
\delta^2_X(t)\defeq\E[X(t)^2] = 2 D t,
\end{equation}
the slope of this MSD being determined by the diffusion coefficient $D$. (ii) The random position $X$ is distributed according to Gaussian statistics
with the probability density
\begin{equation}
\label{eq:Gauss}
p_{X}(x,t)=\f{1}{\sqrt{4\pi D t}}\exp\lt(-\f{x^2}{4 D t}\rt).
\end{equation}
From a random walk perspective these properties emerge from the central limit
theorem for the weakly dependent, identically distributed random variables
\cite{hughes}. Properties \Ref{eq:msd} and \Ref{eq:Gauss} can be readily
obtained from the stochastic equation \cite{langevinEng}
\begin{equation}
\label{eq:LE}
\dot X(t)=V(t),\quad m\dot V(t)=-\lambda V(t)+\sqrt{k_BT \lambda}\xi(t)
\end{equation}
introduced by Langevin, which describes the dynamics of the velocity process $V$
of a particle of mass $m$ in a thermal bath of temperature $T$, where $k_B$ is
Boltzmann's constant and $\lambda$ the damping coefficient. Equation (\ref{eq:LE})
models the interaction of the Brownian particle with the surrounding medium: the
Gaussian white noise term $\sqrt{k_BT\lambda}\xi(t)$ corresponds to the rapid
exchange of momentum between the test particle and the environment. The motion
$X$ is considered slow in comparison to individual bombardments by bath particles.
The term $-\lambda V(t)$ represents the viscosity of the surrounding medium, its
exact magnitude determined by the properties of the liquid and the particle shape,
and thus stands for energy dissipation. Solving the Langevin equation (\ref{eq:LE})
and comparing the stationary value of the mean squared velocity with the thermal
energy due to the equipartition theorem, we obtain the Einstein-Smoluchowski
relation $D=k_BT/(4\lambda m)$ \cite{coffey}.

However, in modern experiments deviations from both of these properties are quite
commonly observed. In particular, anomalous diffusion exhibiting power-law forms
of the MSD,
\begin{equation}
\delta^2_X(t)\sim 2 D t^{\alpha}
\end{equation}
was reported from various physical systems \cite{rwGuide,anTrans,noerregaard,
weissAnDiff}. We distinguish two cases: subdiffusion for $0<\alpha<1$, observed
in the cytoplasm of living biological cells \cite{weakErgJeon,yang,kouExp,
goldingCox,tabei,diRienzo}, various crowded fluids \textit{in vitro}
\cite{szymanski,factNet,banks,lene1} and lipid bilayer membrane systems
\cite{jae-hyung,liBi2pl,akimoto,kneller,matti17,weissAnDiff}; and superdiffusion for
$1<\alpha<2$ wich is related to active biological transport \cite{caspi,reverey,
desposito,goychukmotorpccp} or turbulence \cite{richardson,budaev,liuPlasma,
perriSun}. These anomalous diffusion phenomena cannot be explained solely based
on the Langevin equation \Ref{eq:LE}, which has a too simple structure of memory.
A velocity increment at time $t$ depends exclusively on the present value of
$V(t)$ and the white noise $\xi(t)$, which is independent of the dynamics in
the past. Therefore the future evolution of $V$ is also independent from all but
its most recent value, in other words $V$ is a Markov process \cite{markovB}.

There exist several approaches to anomalous diffusion, which introduce various
degrees of memory \cite{weakErgBreak}. An important extension of the Langevin equation
\Ref{eq:LE}, the generalised Langevin equation (GLE), was promoted in the famous
work of Kubo \cite{kubo} and widely applied in chemical physics \cite{haenggi,
zwanzig}. The GLE is an integro-differential equation of the form
\cite{weiss,kou,sandevGLE,tateishi,goychukGLE,zwanzig}
\begin{equation}
\label{eq:GLE}
m\dot V(t)=-\int_{-\infty}^tV(\tau)K(t-\tau)\dd\tau+\xi(t)
\end{equation}
in which the more complex dependence is reflected both in the memory integral
(of convolution form) with the kernel $K$, and in the stochastic force $\xi$,
which is now described by the covariance function $r_\xi(t)=\mathbb{E}[\xi(
\tau+t)\xi(\tau)]$. For a power-law
kernel the solution of the GLE is an antipersistent motion which models
subdiffusion \cite{weiss,kou,wangGLE} and can be written in terms of a fractional
order Langevin equation \cite{lutz,ergLang,weakErgBreak}.

The GLE has a somewhat special status among stochastic models of anomalous
diffusion, as it can be strictly derived from statistical mechanics. The most
general approach is the projection-operator formalism \cite{zwanzig} but
additional physical insight can be gained from more specific derivations, for
instance, from the Kaz-Zwanzig model of a degree of freedom interacting with
a heat bath of harmonic oscillators \cite{kacGLE,zwanzigGLE}, a test particle
interacting with a continuous field \cite{rey-bellet,pavliotis}, or a Rouse model
describing the conformational dynamics of a monomer in a polymeric bead spring
model of mass points connected by harmonic springs \cite{rouse,panja}. The GLE
with power-law kernel also emerges from a harmonisation of a single file system
of interacting hard core particles \cite{eli_tobias}. It follows from this
derivation that $\xi$ is a stationary Gaussian process, that is, every vector
$[\xi(t_1+\tau), \xi(t_2+\tau),\ldots, \xi(t_n+\tau)]$ has an $n$-dimensional
Gaussian distribution, which does not depend on the time shift $\tau$. Moreover,
the kernel $K$ and the stochastic force are related by the famed Kubo
fluctuation-dissipation theorem $\E[\xi(\tau)\xi(\tau+t)]=\sqrt{k_BT}\times K(t)$
\cite{kubo,knellerFD}. Physically, the GLE with power-law kernel is related to
viscoelastic systems, and was identified as the underlying stochastic process
driving the subdiffusion of submicron tracers in cells, crowded liquids, and
lipid diffusion in simple bilayer membranes \cite{weakErgJeon,reverey,lene1,
szymanski,kneller,jae-hyung}.

\subsection{Non-Gaussian diffusion processes}

However, an additional phenomenon was unveiled in numerous experiments recently.
Namely, not only the assumption of normal diffusion is no longer generally valid,
numerous experiments have shown a new class of diffusive dynamics in which the
fundamental Gaussian property \Ref{eq:Gauss} is violated
\cite{wangNG,metzlerNG,jeonNG, metzlerNG2}; see also the additional references
in \cite{diffDiffChechkin}. In many of these observations the MSD is still
linear, of the form \Ref{eq:msd}, however, the probability density function has
the exponential shape (often called Laplace distribution) \cite{wang2,wangNG,
bhattacharya}
\begin{equation}
\label{eq:LaplPDF}
p_X(x,t)=\f{1}{\sqrt{4D t}}\exp\lt(-\sqrt{\f{1}{Dt}}|x|\rt).
\end{equation}
How can such observations be explained physically? One of the approaches allowing
to explain the emergence of the Laplace distribution is that the measured particle
motion does not correspond to samples of the distribution \Ref{eq:Gauss}, but to a
mixture of individual Gaussian processes with different values of the diffusivities
$D$. In statistics such an object is called a compound or mixture distribution
\cite{mixMod}; in the analysis of diffusion processes this type of model is called
superstatistical \cite{beck2} (which stands for ''superposition of statistics'')
or ''doubly stochastic'' \cite{doubleSts}, which is a term for stochastic models
generalised by replacing some parameter, for instance, $D$, by a random process.
The observations of the Laplace distribution \Ref{eq:LaplPDF} can be justified by
assuming that the diffusion coefficient $D$ is a random variable with exponential
distribution. Every single trajectory is still Gaussian, but the probability
density calculated from the whole ensemble is a compound distribution, in this
case exactly the Laplace distribution \cite{wangNG}.

There are a few physical interpretations that explain the randomness of
$D$. The particles that we observe may not be identical and their different
shapes and interactions with the surroundings could and should affect how
quickly they diffuse. The environment may also be not homogeneous, which
is an expected property of many complex systems, especially biological
ones, such as cell membranes. In this situation the diffusion coefficient
is local and position dependent $D=D(x)$ \cite{posDiff,posDiff2}. If the
particle is moving along trajectory $X(t)$ the effective diffusivity felt
is indirectly time-dependent, $D(t)=D(X(t))$. Time dependence can also be
direct ($D=D(t,X(t))$) if the environment is changing e.g. because of motion
of other particles. Often an approximation is used in which in which $D$ is
assumed to evolve independently from $X$. This is a ''diffusing diffusivity''
approach proposed by Chubinsky and Slater \cite{chubynsky}, which is currently
being actively developed \cite{diffDiffCherstvy,diffDiffJain,diffDiffChechkin}.

According to the superstatistics approach \cite{beck2} the different diffusivities
correspond to the motion of one given particle in a region with a given $D$-value.
At sufficiently short time scales the observed particles are relatively localised
in such a region, and $D$ can be considered to be constant for each trajectory. The
whole ensemble of particles behaves as a system with random $D$ with a distribution
of $D$-values mirroring the spacial and temporal dispersion of $D(t,x)$ \cite{manzo}.
Beck proposed that in turbulent media one could consider the Langevin equation
\Ref{eq:LE} to be valid, however, the effective temperature is random such that
$(k_BT)^{-1}$ is distributed according to $\chi^2$ statistics \cite{beck,beck2,
beck3}.

We note here that viscoelastic anomalous diffusion with Laplace shape of the
probability density function was observed for the motion of messenger RNA
molecules in the cytoplasm of bacteria and yeast cells \cite{lampo,ralf_lampo},
while stretched Gaussian shapes were unveiled in the motion of lipids in
protein-crowded lipid biliayer systems \cite{jeon_prx}.

In what follows we study a natural extension of this idea: what if not the
temperature, but the properties of the stochastic force in equations \Ref{eq:LE}
and \Ref{eq:GLE} is random? Such an assumption may be justified in the same way
as the randomness of $D$. Namely, this situation can be realised in an ensemble
of particles with varying systems parameters, or in non-homogenous media. This
approach resembles to some degree models such as the GLE with kernels, that are
a mixture of more elementary functions \cite{sandev, tateishi}. Similar ideas
appear in financial modelling with ''gamma-mixed Ornstein-Uhlenbeck process''
\cite{mixOU}. However in these models the observed trajectories are not
examples of compound distributions but result from deterministic dynamics,
which can be interpreted as an average over random local dynamical laws. In
our approach the studied processes are truly superstatistical.

The paper is structured as follows. In section \ref{s:ssGLE} we introduce the
GLE with random parameters and discuss its elementary properties. Section
\ref{s:cOU} then considers the concrete case of a compound Ornstein-Uhlenbeck
process within the superstatistical approach. In section \ref{morecomplex}
more complex types of memory are proposed and studied, including oscillatory
regimes. Our findings are discussed in section \ref{summary}. In the appendix
some more technical details are presented.

\section{Generalised Langevin equation with random parameters}
\label{s:ssGLE}

Our starting point is the GLE assumed to depend on some parameter $c$, which
may describe the type of diffusing particle and/or the local properties of its
environment. The parameter $c$ can, in principle, be a number or a vector. In
the GLE the stochastic force then also becomes parametrised by $c$, $\xi=\xi_c$. Due
to their coupling via the Kubo fluctuation-dissipation relation, also the memory
kernel depends on $c$, $K=K_c$ (see below). The solution of the GLE, the velocity
and position processes can then also be considered to be functions of this
parameter, $V=V_c$, $X=X_c$. These phase space coordinates thus solve the set
of equations
\begin{align}
\label{eq:ssGLE}
\nonumber
m\dot V_c &= -\int_{-\infty}^tV_c(\tau)K_c(-t-\tau)\dd\tau+\xi_c,\\
\dot X_c &= V_c.
\end{align}
The constant $\sqrt{k_BT}$ and the mass $m$ actually only rescale the solutions.
Considering $c$ dependent mass $m$ and temperature $T$ would result in a $c$
dependent diffusion constant $D_c$. This type of influence was extensively
studied before \cite{beck,beck2,beck3,touchette,straeten}, and therefore we will
omit this ramification in our analysis and assume $m=k_BT=1$ in what follows.

In the above definition we tacitly assumed that the introduction of the parameter
$c$ does not change the spatially local structure of the GLE, and we assume that
the fluctuation-dissipation theorem remains valid in the form
\begin{equation}
\mathbb{E}[\xi_c(\tau)\xi_c(\tau+t)]=\sqrt{k_B T} K_c(t).
\end{equation}
By design the GLE is a spatially local equation. The variable $X$ interacts with
the heat bath only in its neighbourhood. The bath degrees of freedom themselves
do not interact with each other directly, which prohibits spatial long-range
correlations. Long-time correlations can still be present, but they result from
the interactions between $X$ and the bath degrees of freedom, which ''store'' the
memory structure for a long time, but do so only locally. That means for each
fixed value $c$ the fluctuation-dissipation theorem should still hold.

For every $c$, the GLE can be solved using the Green's function formalism.
The stationary solution of equation \Ref{eq:ssGLE} is given by 
\begin{equation}
V_c(t)=\int_{-\infty}^t\xi_c(\tau)G_c(t-\tau)\dd\tau,
\end{equation}
where the Green's function $G$ solves the equation
\begin{equation}
\label{eq:defG}
\dot G_c(t)=-\int_0^t G_c(\tau)K_c(t-\tau)\dd\tau+\delta(t).
\end{equation}
Equivalently, $G_c$ is the inverse Laplace transform of
\begin{equation}
\widetilde{G_c}(s)=\f{1}{s+\widetilde{K_c}(s)},
\end{equation}
where by $\widetilde{G_c}(s)$ we denote the Laplace transform of $G_c(t)$.

Generally the superstatistical solution $V_C$ and $X_C$ of the GLE emerges when the parameter $c$ is drawn from some distribution, which we denote by substituting a big letter $C$ for it. In order to get a better
feeling, as a guiding example let us consider the simple case of a discrete set
of local environments or types of particles. We number them by $c=1,2,3,\ldots$
The random variable $C$ with distribution $\pr(C=k)=p_k$ then describes how many
trajectories are evolving in each environment or correspond to each particle type.
With probability $p_k$ the observed trajectory evolved according to the GLE
\Ref{eq:ssGLE} with kernel $K_k$ and stochastic force $\xi_k$. In general $C$ may
not be discrete, but can be a continuous parameter. The latter case is more complex
and interesting, allowing to model a wider range of phenomena, and this will be our
main point of interest in the following.

When we consider the solutions of the superstatistical GLE (that is equation \Ref{eq:ssGLE} together with a model distribution for $C$) we are
interested in two types of observables: ensemble and time averages. Ensemble
averages correspond to quantities relevant when an experiment averages over
many particles, as in the pioneering experiments of Perrin \cite{perrin1,perrin2}.
Single particle tracking experiments with sufficiently long individual particle
traces, as those introduced by Nordlund \cite{nordlund}, are typically evaluated
by time averages \cite{physicstoday}. For the case of ensemble averages the
situation is simple, these quantities can be calculated using the so-called
''tower property'', which can be applied to any random function $f(X)$
\begin{equation}
\E_X[f(X)]=\E_C[\E_{X|C}[f(X)|C]]=\int\dd \pr_C(c)\int\dd \pr_{X|c}(x)\ f(x).
\end{equation}
In what follows we omit the subscripts in the notation $\E_{(\bd\cdot)}$ and
use the convention that the variables will always be averaged with respect to
their natural distribution. In order to calculate ensemble averages of $V_C$
and $X_C$ we simply need to calculate ensemble averages of $V_c,X_c$ for fixed
$c$ and average them over the distribution of $C$. In particular, if $C$ and
$X_c$ have the probability density (PDF) $p_C$ and $p_{X_c}$, the density of
$X_C$ becomes
\begin{equation}
p_{X_C}(x,t)=\E\lt[p_{X_C}(x,t|C)\rt]=\int p_C(c)p_{X_c}(x,t)\dd c.
\end{equation}

For the time averages, the most commonly used quantity is the time averaged MSD,
which for a trajectory of length $\mathcal T$ (\emph{observation time\/}) reads
\cite{physicstoday,noerregaard,weakErgBreak}
\begin{eqnarray}
\nonumber
\overline{\delta^2(t;\mathcal T)}&\defeq &\f{1}{\mathcal T-t}\int_0^{\mathcal T-t}
\big(X_C(\tau+t)-X_C(\tau)\big)^2\dd\tau\\
&=&\f{1}{\mathcal T-t}\int_0^{\mathcal T-t}\lt(\int_{\tau}^{\tau+t}V_C(\tau')
\dd\tau'\rt)^2\dd\tau.
\end{eqnarray}
The last form stresses that $\overline{\delta^2(t;\mathcal T)}$ can be viewed
as a function of the velocity $V_C$, which here is a stationary process. In
this work we will be mostly interested in the limit $\mathcal T\to\infty$
for which one can omit subtracting $t$ and use the simpler average $\lim_{\mathcal T\to\infty}\f{1}{
\mathcal T}\int_0^{\mathcal T}(\bd\cdot)\dd\tau$
and determine it using tools from ergodic theory. For every choice of $c$
the stochastic force $\xi_c$ is stationary and Gaussian, and its covariance
function decays to zero, $r_{\xi_c}(t)\to 0$ as $t\to\infty$. The famous Maryuama
theorem \cite{maruyama,dym} guarantees that in such a case the process is
mixing, in particular, it is ergodic. So the stationary solution $V_c$ of the
GLE \Ref{eq:ssGLE} must be mixing and ergodic, as well, for every choice of $c$
\cite{ergSpec}. Any time average coincides with the ensemble average, that is,
for every function of state $f$\footnote{Note, however, that generally a process
described by the GLE can be \emph{transiently\/} non-ergodic and ageing, as shown
in references \cite{lene1,jeon_pre12,kursawe}.}
\begin{equation}
\label{eq:ergTh}
\overline{f(V_c)}\defeq\lim_{\mathcal T\to\infty}\f{1}{\mathcal T}\int_0^{\mathcal T}
f(V_c(\tau))\dd\tau=\E[f(V_c(0))].
\end{equation}
However, the superstatistical solution $V_C$ cannot be ergodic as averaging
over one trajectory one cannot gain insight into the distribution of $C$. But,
the process $V_C$ is still stationary, as, for every $c$, $V_c$ is stationary.
In such a case the behaviour of the time averages is determined by Birkhoff's
theorem \cite{dym,doob} which guarantees that
\begin{equation}
\overline{f(V_C)}=\lim_{\mathcal T\to\infty}\f{1}{\mathcal T}\int_0^{\mathcal T}
f(V_C(\tau))\dd\tau=\E[f(V_C(0))|\mathcal M].
\end{equation}
All time averages converge to a random variable $\E[f(V_C(0))|\mathcal M]$,
which is an expected value conditioned by $\mathcal M$ summarising all constants
of motion of $V_C$. For isolated systems $\mathcal M$ would correspond to the
energy, total momentum and similar quantities. In our case here every trajectory
$V_c$ itself is ergodic, so it has no internal constants of motion. Therefore the
only constants of motion are the local states of the environment, denoted by $C$.
This statement is intuitively reasonable: given a trajectory evolving with $C=c$
all time averaged statistics converge to the values corresponding to the solution
of the GLE with $K_c$ and $\xi_c$, which, due to the ergodic theorem \Ref{eq:ergTh}
are exactly the conditional expected values $\E[f(V_C(0))|C=c]=\E[f(V_c(0))]$. For the MSD this
means that 
\begin{equation}
\overline{\delta^2(t)}\defeq\lim_{\mathcal T\to\infty}\overline{\delta^2(t;\mathcal T)}=\E[X_C(t)^2|C]=\delta^2_{X_C}(t|C),
\end{equation}

which is a function of the random parameter $C$ and time $t$. One consequence of
this is that the ergodicity breaking parameter \cite{he,ergPar,weakErgBreak} never
equals zero and does not converge to zero even in the asymptotic limit $t\to\infty$,
\begin{equation}
\label{eq:EB}
\text{EB}(t)\defeq\f{\E\lt[\overline{\delta^2(t)}^2\rt]}{\E\lt[\overline{
\delta^2(t)}\rt]^2}-1=\f{\E\lt[\big(\delta^2_{X_C}(t|C)\big)^2\rt]}{\E\lt[\delta
^2_{X_C}(t|C)\rt]^2}-1\neq 0,
\end{equation}
for any $t\neq0$, as the mean of a square equals the square of a mean only for
non-random variables.

The ensemble averaged MSD can be directly obtained from the Green's function
$G_c(t)$. Namely, proposition \ref{prp:cov} in the Appendix proves that the
covariance function of $V_c$ equals the Green's function, $r_{V_c}=
G_c$\footnote{Without the rescaling $m=k_BT=1$, $r_{V_c}$ it equals $k_BTm^{-1}
G_c$}
such that the ensemble averaged MSD of $X_c$ reads
\begin{equation}
\label{eq:msdGen}
\delta^2_{X_c}(t)=2\int_0^t\dd\tau_1\int_0^{\tau_1}\dd\tau_2\ G_c(\tau_2).
\end{equation}
Moreover, as $G_c$ is a covariance function, it is bounded, $|G_c(0)|\le G_c(0)=
1$. Thus, from relation \Ref{eq:msdGen} we see that $\delta^2_{X_c}(t)\le t^2/2$,
so the MSD is always finite and the motion governed by the superstatistical GLE
is sub-ballistic.
The result for the MSD assumes a particularly simple form in Laplace space,
\begin{equation}
\label{eq:msdLapl}
\widetilde{\delta^{2}_{X_C}}(s)=2s^{-2}\E\lt[\widetilde{G_C}(s)\rt]=2s^{-2}\widetilde{
r_{V_C}}(s).
\end{equation}
Note that $G_c(0)=1$ also implies that $\delta^2_{V_c}(t)=1$. For any $t$
the value $V_C(t)$ is not superstatistical, it is simply a Gaussian variable
with variance $1$, which is the same as for $V_c(t)$ with any $c$. At
the same time the covariance function $r_{V_C}$ is decaying as a mixture
of decaying functions $r_{V_c}$. Without careful consideration this may seem
contradictory: the Maryuama theorem states that if a  stationary Gaussian
process has a decaying covariance function, it is mixing and ergodic, but
$V_C$ is stationary, Gaussian at every $t$, has a decaying covariance function
and is not ergodic.

The solution to this seeming contradiction is the fact that while $V_C$ is
Gaussian at every instant of time $t$, it is itself not a Gaussian process.
For a stochastic process to be Gaussian, it is not sufficient that is has a
Gaussian marginal distribution but also a Gaussian joint distribution. The
solution $V_C$ is an interesting physical example of an object witch has
Gaussian marginals, but non-Gaussian memory structure. Such processes are
well-known to exist: it is enough to take some non-Gaussian process $X(t)$ and
transform it using its own cumulative distribution function, $Y(t)=F_t(X(t)),
F_t(x)=\pr(X(t)<x)$. The resulting process $Y(t)$ has uniform distribution for
every $t$, and it is enough to transform it a second time using a normal quantile
function to obtain a process with Gaussian PDF, yet a complicated and
particularly non-Gaussian type of dependence. However this construction can be
considered artificial and without physical meaning. The unusual non-Gaussianity
of $V_C$ here arises naturally from the physical model. Process $V_C$ could be
very misleading during the analysis of measured data, using only basic statistical
methods it will seem Gaussian. We will show techniques which can be used to unveil
its non-Gaussianity in the next Section for specific examples.

\section{Compound Ornstein-Uhlenbeck process}
\label{s:cOU}

\subsection{Overview of the model}

\begin{figure}
\centering
\includegraphics[width=12cm]{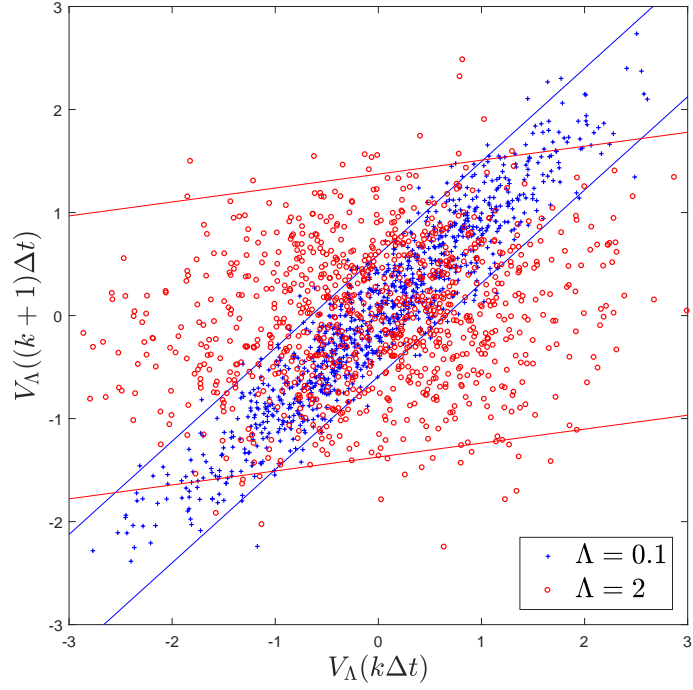}
\caption{Phase plot of the compound Ornstein-Uhlenbeck process with $\pr(\Lambda=
1/10)=\pr(\Lambda=2)=1/2$. We took $\Delta t=1$. Solid lines correspond to 95\%
conditional quantiles of the noise $Z_k$ in both populations.}
\label{f:plotAR}
\end{figure}

The classical Langevin equation can be considered as an approximation of the GLE
in which the covariance function $r_\xi$ decays very rapidly in the relevant
time scale. The solution of the Langevin equation exhibits many properties typical
to the GLE in general. We fix the mass of the particle and the bath temperature,
so equation \Ref{eq:LE} is governed solely by the parameter $\lambda$. The
superstatistical solution is thus $V_\Lambda$, where $\Lambda>0$ is a random
variable, which can be interpreted as a local viscosity value. The Langevin
equation can be solved using the integrating factor $\exp(\Lambda t)$, which
yields the stationary solution
\begin{equation}
V_\Lambda(t)=\sqrt{\Lambda}\int_{-\infty}^t\xi(\tau)\e^{-\Lambda(t-\tau)}\dd\tau.
\end{equation}
The solution $V_\lambda$ for fixed $\lambda$ is often called Ornstein-Uhlenbeck
process, so $V_\Lambda$ may be called a compound Ornstein-Uhlenbeck process. It
can also be represented in Fourier space. Calculating the Fourier transform of
equation \Ref{eq:LE} demonstrates that
\begin{equation}
V_\Lambda(t)=\int_{-\infty}^\infty\widehat\xi(\omega)\f{\sqrt{\Lambda}}{\I\omega+\Lambda}
\e^{\I\omega t}\dd\omega,
\end{equation}
where we note that the Fourier transform of Gaussian white noise $\widehat\xi$
is another Gaussian white noise. Another useful representation is the recursive
formula, which is fulfilled by process in discretised time moments. If we solve
the Langevin equation \Ref{eq:LE} using the integrating factor $\exp(\Lambda t)$
but integrate from time $k\Delta t$ to $(k+1)\Delta t$ we obtain
\begin{align}
&V_\Lambda((k+1)\Delta t)=\e^{-\Delta t \Lambda} V_\Lambda(k\Delta t)+Z_k
\nonumber\\
&Z_k=\sqrt{\Lambda}\int_{k\Delta t}^{(k+1)\Delta t}\xi(\tau) \e^{-\Lambda((k+1)\Delta t-\tau)}
\dd\tau\deq \f{1}{\sqrt{2}}\sqrt{1-\e^{-2\Delta t \Lambda}}\xi_k,
\end{align}
where the noise $Z_k$ has the same distribution as a Gaussian discrete white
noise $\xi_k$ multiplied by a random constant. The series $Z_k$ is, conditionally
on $\Lambda$, independent from past values $V_\Lambda(j\Delta t)$, with $j< k$.
Such a process is called a random-coefficient autoregressive process of order 1,
in short AR(1) \cite{BD,BJ} with autoregressive coefficient $\exp(-\Delta t
\Lambda)$.  When there are only few distinct populations and $\Lambda$
has only few possible values, they can even be recognised on the phase plot of
$y=V_\Lambda((k+1)\Delta t)$ versus $x=V_\Lambda(k\Delta t)$, see Figure
\ref{f:plotAR}. There, the two distinct populations with different autoregressive
coefficients can be distinguished. Both have Gaussian distribution, but each one
has a distinct elliptical shape. The total distribution, as a mixture of two
ellipsoids, is not Gaussian, nor even elliptical. The projection of the joint
distribution on $x$ or $y$ axis are the PDF of $V_\Lambda(t)$ and are Gaussian,
thus, one needs at least a two-dimensional phase plot to reveal the non-Gaussianity
of $V_\Lambda$. For a larger number of populations the phase plot would be much
less clear, but the huge advantage of this method is that it works even for
trajectories of very short length.

The situation becomes more complex and interesting when $\Lambda$ assumes a
continuous distribution. The covariance function of the Ornstein-Uhlenbeck
process is
\begin{equation}
r_{V_\lambda}(t)=\f{1}{2}\e^{-\lambda t}.
\end{equation}
Here some care needs to be taken, as the factor $1/2$ differs from the covariance
of the solution of the GLE for which generally $r_V(0)=1$ (this is due to the fact
that it corresponds to a degenerate Dirac delta kernel). If $\Lambda$ has the PDF
$p_\Lambda$, the covariance function of $V_\Lambda$ is
\begin{equation}
r_{V_\Lambda}(t)=\f{1}{2}\int_0^\infty p_\Lambda(\lambda) \e^{-\lambda t}\dd\lambda,
\end{equation}
so it is the Laplace transform of $p_\Lambda$: in probabilistic language this
quantity would be called a moment generating function of the variable $-\Lambda$.
For instance, if $\Lambda$ is a stable subordinator with index $0<\alpha<1$
\cite{taqqu} the covariance function is the stretched exponent
\begin{equation}
r_{V_\Lambda}(t)=\f{1}{2}\e^{-\sigma_\alpha t^\alpha},
\end{equation}
which is a common relaxation model \cite{kakalios,piazza,kopf,weronSExp},
sometimes referred to as Kohlrausch-Williams-Watts relaxation \cite{kohlrausch,
watts}.

If $\Lambda$ can be decomposed into a sum of two independent random variables
$\Lambda=\Lambda_1+\Lambda_2$, the corresponding covariance function is a product,
\begin{equation}
\label{eq:cOUmultipl}
r_{V_\Lambda}(t)=2r_{V_{\Lambda_1}}(t)r_{V_{\Lambda_2}}(t).
\end{equation}
Therefore, in this model various kinds of truncations of the kernel correspond
to a decomposition of $\Lambda$, for instance, if $\Lambda=\lambda+\Lambda'$
with deterministic $\lambda>0$, the covariance function $r_{V_\Lambda}$ will be
truncated by $\exp(-\lambda t)$.

Some general observations about the behaviour of $r_{V_\Lambda}$ can be made.
When $\Lambda$ has a distribution supported on an interval, such as $\lambda_1
<\Lambda<\lambda_2$, and its PDF has no singularity, it is necessarily bounded,
that is, $m\le p_\Lambda(\lambda)\le M$. In this case,
\begin{equation}
\label{eq:LbdUnif}
\f{m}{2}\int_{\lambda_1}^{\lambda_2}\e^{-\lambda t}\dd\lambda\le r_{V_\Lambda}(t)
\le\f{M}{2}\int_{\lambda_1}^{\lambda_2}\e^{-\lambda t}\dd\lambda.
\end{equation}
The integrals on the left and right have asymptotics of the form
\begin{equation}
\int_{\lambda_1}^{\lambda_2}\e^{-\lambda t}\dd\lambda =\f{1}{t}\lt(\e^{-\lambda_1
t}-\e^{-\lambda_2 t}\rt)\sim\f{1}{t}\e^{-\lambda_1 t}.
\end{equation}
Here we introduce the notation of an asymptotic inequality, which will be
useful later on. We write $f\lesssim g$ if $f\sim h\le g$ for some function
$h$.\footnote{This is similar to the Landau-$\mathcal O$ notation, but when
we write $\lesssim$ we also include the value of the constant factor which would be
omitted writing $r_{V_\Lambda}=\mathcal O(t^{-1}\e^{-\lambda_1 t})$.}
Using this notion we can write the above results as
\begin{equation}
\label{eq:cOUBoundedAs}
\f{m}{2t}\e^{-\lambda_1 t}\lesssim r_{V_\Lambda}(t)\lesssim \f{M}{2t}
\e^{-\lambda_1 t}.
\end{equation}
This equation proves that $c_1 t^{-1}\e^{-\lambda_1 t}\lesssim r_{V_\Lambda}(t)
\lesssim c_2 t^{-1}\e^{-\lambda_1 t}$ for some constants $0<c_1<c_2$. Further on
we will denote this property by $r_{V_\Lambda}\asymp t^{-1}\e^{-\lambda_1
t}$.\footnote{The same situation is sometimes denoted in terms of the
''large theta'' notation, that is $r_{V_\Lambda}=\Theta(t^{-1}\e^{-\lambda_1
t})$.} When $\Lambda$ is distributed uniformly, $m=M=(\lambda_2-\lambda_1)^{-1}$,
and the asymptotic $\asymp$ becomes stronger, that is, $r_{V_\Lambda}(t)\sim
(2(\lambda_2-\lambda_1)t)^{-1}\e^{-\lambda_1 t}$. This distribution of
$\Lambda$ is important from a practical standpoint, because it is a maximal
entropy distribution supported on the interval $[\lambda_1,\lambda_2]$, so it can
be interpreted as the choice taken using the weakest possible assumptions.

Heavier tails of $r_{V_\Lambda}$ may be observed only when the distribution of
$\Lambda$ is concentrated around $0^+$.The most significant case of such a
distribution is a power law of the form $p_\Lambda(\lambda)\sim \lambda^{\alpha-1}$,
with $\lambda\to 0^+$ and $\alpha>0$. For any distribution of this type Tauberian
theorems guarantee that the tail of the covariance has a power law tail
\cite{feller,taub}
\begin{equation}
r_{V_\Lambda}(t)\sim\f{\Gamma(\alpha)}{2} t^{-\alpha},\quad t\to\infty.
\label{eq:cOUPowerLaw}
\end{equation}
For $\alpha<1$ the process $V_\Lambda$ exhibits a long memory. This observation
can be refined as follows. In Proposition \ref{prp:slowVar} i) we present
a generalised Tauberian theorem, which states that if the PDF of $\Lambda$
contains slowly-varying factor $L$, then the tail of the covariance contains
the factor $L(t^{-1})$. One example of such a slowly-varying factor is $|\ln(
\lambda)|^\beta,\beta>0$, so heavy tails of the covariance of the power law form
$t^{-\alpha}\ln(t)^\beta$ can also be present for compound Ornstein-Uhlenbeck
process if the distribution of $\Lambda$ exhibits a logarithmic behaviour at
$0^+$. This observation proves that this equation can also describe ultra-slow
diffusion and can be considered as an alternative to more complex models
based on distributed order fractional derivatives \cite{eab,sandev}.

In Section \ref{s:ssGLE} we noted that the superstatistical solutions of the
Langevin equations are not Gaussian, however, they can be easily mistaken to
be Gaussian. The marginal distributions of $V_\Lambda$ are Gaussian at any time
$t$, only the joint distributions are not. This means that the multidimensional
PDF of the variables $V_\Lambda(t_1),V_\Lambda(t_2),\ldots, V_\Lambda(t_n)$ does
not have Gaussian shape. This fact is easy to observe studying the characteristic
function of the two point distribution, which is a Fourier transform of the
two-point PDF. Let us fix $V_\Lambda(\tau)$, $V_\Lambda(\tau+t)$ and define
the two-point characteristic function as
\begin{equation}
\phi_\Lambda(\bd\theta,t)\equiv\E\lt[\e^{\I(\theta_1 V_\Lambda(\tau)+\theta_2
V_\Lambda(\tau+ t))}\rt],\quad \bd\theta=[\theta_1,\theta_2].
\end{equation}
For any deterministic $\lambda$ this function is determined by the covariance
matrix $\Sigma_t$ of the pair $V_\lambda(\tau),V_\lambda(\tau+t)$,
\begin{equation}
\phi_\lambda(\bd\theta,t)=\e^{-\f{1}{2}\bd\theta^{\mathrm T}\Sigma_t\bd\theta},
\quad \Sigma_t=\f{1}{2}\bgmx 1, &\e^{-\lambda t} \\ \e^{-\lambda t}, & 1\emx,
\end{equation}
so in the superstatistical case the characteristic function reads
\begin{equation}
\label{eq:phi}
\phi_\Lambda(\bd\theta,t)=\E[\phi_\Lambda(\bd\theta,t|\Lambda)]=\e^{-\f{1}{4}
\theta_1^2}\e^{-\f{1}{4}\theta_2^2}\E\lt[\e^{-\f{1}{2}\theta_1\theta_2\e^{
-\Lambda t}}\rt].
\end{equation}
As we argued, the marginal factors $\exp(-\theta_1^2/4),\exp(-\theta_2^2/4)$ are
indeed Gaussian, but the cross factor describing the interdependence is not. The
function $\phi_\Lambda$ would describe a Gaussian distribution if and only if the
factor $\E\lt[\exp(\theta_1\theta_2\e^{-\Lambda t}/2)\rt]$ had the form $\exp(a
\theta_1\theta_2)$. But we see that it is in fact a moment generating function
of the variable $\exp(-\Lambda t)$ at point $\theta_1\theta_2/2$, which is an
exponential if and only if $\Lambda$ equals one fixed value with probability
unity. The compound Ornstein-Uhlenbeck is \emph{never Gaussian\/} for
non-deterministic $\Lambda$.

This property is also evident if we calculate the conditional MSD of $X_\Lambda$,
\begin{equation}
\label{eq:condmsd}
\delta^2_{X_\Lambda}(t|\Lambda)=\f{1}{2\Lambda}t+\f{1}{2\Lambda^2}\lt(\e^{
-\Lambda t}-1\rt).
\end{equation}
At short times $t$ this approximately is $t^2/4$, so the distribution is nearly
Gaussian and the motion is ballistic. However at long $t$ the dominating
term is $(2\Lambda)^{-1}t$, so we see that if $\E[\Lambda^{-1}]<\infty$, the
integrated compound Ornstein-Uhlenbeck process describes normal diffusion
with random diffusion coefficient $D=(4\Lambda)^{-1}$. Such a situation
occurs when the distribution $\Lambda$ is not highly concentrated around
$0^+$. When $\Lambda$ has a power-law singularity as in \Ref{eq:cOUPowerLaw},
that is $\lambda^{\alpha-1}$ at $0^+$, $0<\alpha<1$, this condition is not
fulfilled: $\E[\Lambda^{-1}]=\infty$. But in this situation the assumptions
required for the Tauberian theorem hold and we can apply it twice: first
for relation \Ref{eq:cOUPowerLaw}, to show that $\widetilde{r_{V_\Lambda}}(s)\sim
2^{-1}\Gamma(\alpha)\Gamma(1-\alpha) s^{\alpha-1},s\to0^+$ and the second
time for relation \Ref{eq:msdLapl}, to prove that
\begin{equation}
\delta_{X_\Lambda}^2(t)\sim\f{2\Gamma(\alpha)}{(1-\alpha)(2-\alpha)} t^{2-\alpha},
\quad t\to\infty.
\end{equation} 
In this regime the system is superdiffusive. The transition from superdiffusion
($0<\alpha<1$) to normal diffusion ($1\le\alpha$) is unusual among diffusion
models. Fractional Brownian motion and fractional Langevin equation
\cite{lutz,burov,kou} undergo transitions from super- to subdiffusion at a
critical point of the control parameter. This is so as in these models in this
the change of the diffusion type is caused by the change of the memory type from
persistent to antipersistent. But the Ornstein-Uhlenbeck process models only
persistent dependence, so the mixture of such motions also inherits this
property. For $1\le \alpha$ (and any other case when $\E[\Lambda^{-1}]<\infty$)
this dependence is weak enough for the process to be normally diffusive, for
smaller values of $\alpha$ it induces superdiffusion.

In the introduction we already mentioned that it is commonly observed that the
distribution of the position process is double exponential, see equation
\Ref{eq:LaplPDF} and references below. This exact distribution is observed when
$D$ has an exponential distribution $\mathcal E(\beta)$ with PDF
\begin{equation}
p_D(d)=\beta\e^{-\beta d}.
\end{equation}
For the corresponding compound Ornstein-Uhlenbeck process the distribution of
$\Lambda$ is given by $\Lambda=(4D)^{-1}$ and for such a choice the process
models normal diffusion with a Laplace PDF. Moreover the covariance function
of the velocity process is
\begin{align}
\label{eq:covLaplpdf}
\nonumber
r_{V_{(4D)^{-1}}}(t)&=\f{1}{2}\E\lt[\e^{-\f{t}{4D}}\rt]\\
\nonumber
&=\f{\beta}{2}\int_0^\infty\e^{-\f{t}{4d}}\e^{-\beta d}\dd d\\
\nonumber
&\overset{\f{t}{4d}\to d}{=}\f{\beta t}{8}\int_0^\infty\e^{-d-\beta t(4d)^{-1}}
\f{1}{d^2}\dd d\\
&=\f{\sqrt{\beta t}}{2}K_1(\sqrt{\beta t}),
\end{align}
where we used one of the integral representations of the modified Bessel function
of the second kind $K_1$ (see \cite{DLMF}, formula 10.32.10). This function has
the asymptotic $K_1(z)\sim \sqrt{\pi/2}\exp(-z)z^{-1/2}$, $z\to\infty$ (\cite{DLMF},
formula 10.40.2), so the covariance function behaves like
\begin{equation}
\label{eq:laplA}
r_{V_{(4D)^{-1}}}(t)\sim \sqrt{\f{\pi}{8}}(\beta t)^{1/4}\e^{-\sqrt{\beta t}},
\quad t\to\infty.
\end{equation}

\begin{figure}
\centering
\includegraphics[width=16cm]{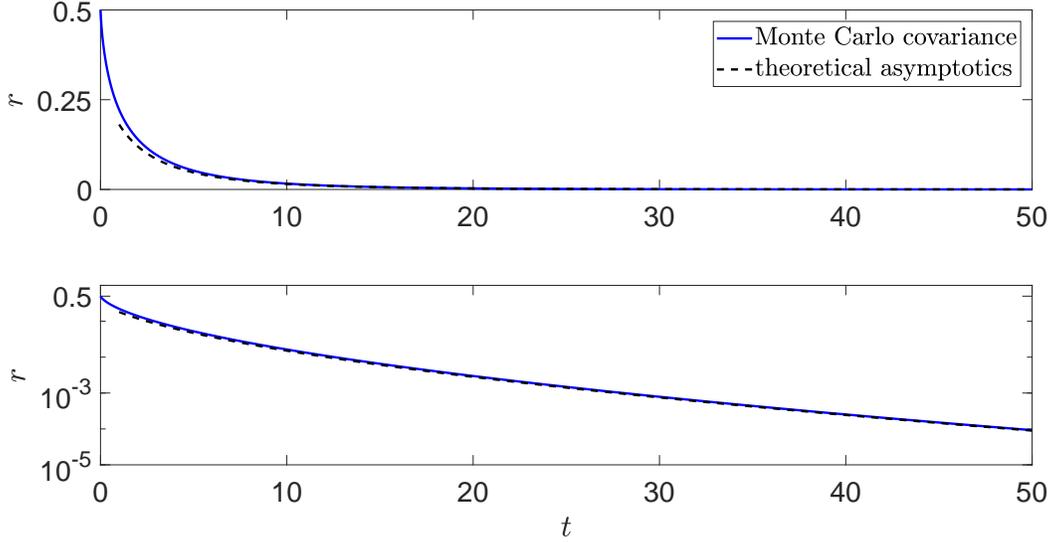}
\caption{Covariance function from Monte Carlo simulations for a system with
Laplace PDF of the position process, together with the theoretical asymptotics.
The sample size is $10^6$, $(2\Lambda)^{-1}\deq \mathcal E(1)$. The covariance
is shown in normal and semi-logarithmic scale. The convergence to the predicted
asymptotic behaviour is excellent. The full solution \Ref{eq:covLaplpdf} is not
shown, it fully overlaps with the simulations results.}
\label{f:laplCov}
\end{figure}

This behaviour is shown in Figure \ref{f:laplCov}, where we present the covariance
function corresponding to the Laplace distributed $X_\Lambda(t)$ with random
diffusion coefficient $D\deq \mathcal E(2)$. We do not present the Bessel function
\Ref{eq:covLaplpdf}, as it appears to be indistinguishable from the result of
the Monte Carlo simulation. Figure \ref{f:laplCov} also shows how to distinguish
this behaviour from an exponential decay on a semi-logarithmic scale: the
covariance function and its asymptotic are concave, which is mostly visible for
short times $t$.

Analysing the shape of the covariance function can serve as a method to
distinguish between a superstatistic introduced by a local effective
temperature and the distribution of mass from superstatistics caused by the
randomness of the viscosity $\Lambda$. In the former case the resulting decay
is exponential (as in the non-superstatistical Langevin equation) or even zero
for a free Brownian particle, in the latter case it is given by relation
\Ref{eq:covLaplpdf}.

\subsection{Gamma distributed $\Lambda$}

In order to better understand the superstatistical Langevin equation we will
consider a simple model with one particular choice for the distribution
$\Lambda$. After going through this explicit example we will come back to
the general case at the end of this section.

A generic choice for the $\Lambda$ distribution is the Gamma distribution
$\mathcal G(\alpha,\beta)$ with the PDF
\begin{equation}
\label{eq:gammaDistr}
p_\Lambda(\lambda)=\f{\beta^\alpha}{\Gamma(\alpha)}\lambda^{\alpha-1}\e^{-\beta
\lambda}, \quad \alpha,\beta>0.
\end{equation}
This corresponds to a power law at $0^+$ which is truncated by an exponential.
As the conditional covariance function is an exponential, too, many integrals
which in general would be hard to calculate, in this present case turn out to
be surprisingly simple.

The Gamma distribution is also a convenient choice because many of its special
cases are well established in physics. The Erlang distribution is the special
case of expression \Ref{eq:gammaDistr} when $\alpha$ is a natural number. An
Erlang variable with $\alpha=k$ and $\beta$ can be represented as the sum of
$k$ independent exponential variables $\mathcal E(\beta)$, in particular, for
$k=1$ it is the exponential distribution itself. The Chi-square distribution
$\chi^2(k)$ is also a special case of expression \Ref{eq:gammaDistr} where
$\alpha=k/2,\beta = 1/2$. The Maxwell-Boltzmann distribution corresponds to the square root of
$\chi^2(3)$, and the Rayleigh distribution to the square root of $\chi^2(2)$.

We already know from relation \Ref{eq:cOUPowerLaw} that $r_{V_\Lambda}$ has a
power tail $\sim 2^{-1}\beta^\alpha t^{-\alpha}$, more specifically, direct
integration yields
\begin{equation}
\label{eq:rcOU}
r_{V_\Lambda}(t)=\f{1}{2}\f{1}{\lt(1+t/\beta\rt)^\alpha}.
\end{equation}
This is solely a function of the ratio $t/\beta$ which suggests that the parameter
$\beta$ changes the time scale of the process. Indeed, for any $\lambda$ the
process $V_\lambda(bt)$ is equivalent to $V_{b\lambda}(t)$, because the
Gaussian process is determined by its covariance function, which in both
cases is the same. Therefore, also the compound process $V_\Lambda(bt)$ is
equivalent to $V_{b\Lambda}(t)$ and $b\Lambda$ has the distribution $\mathcal
G(\alpha,\beta/b)$.

\begin{figure}
\centering
\includegraphics[width=16cm]{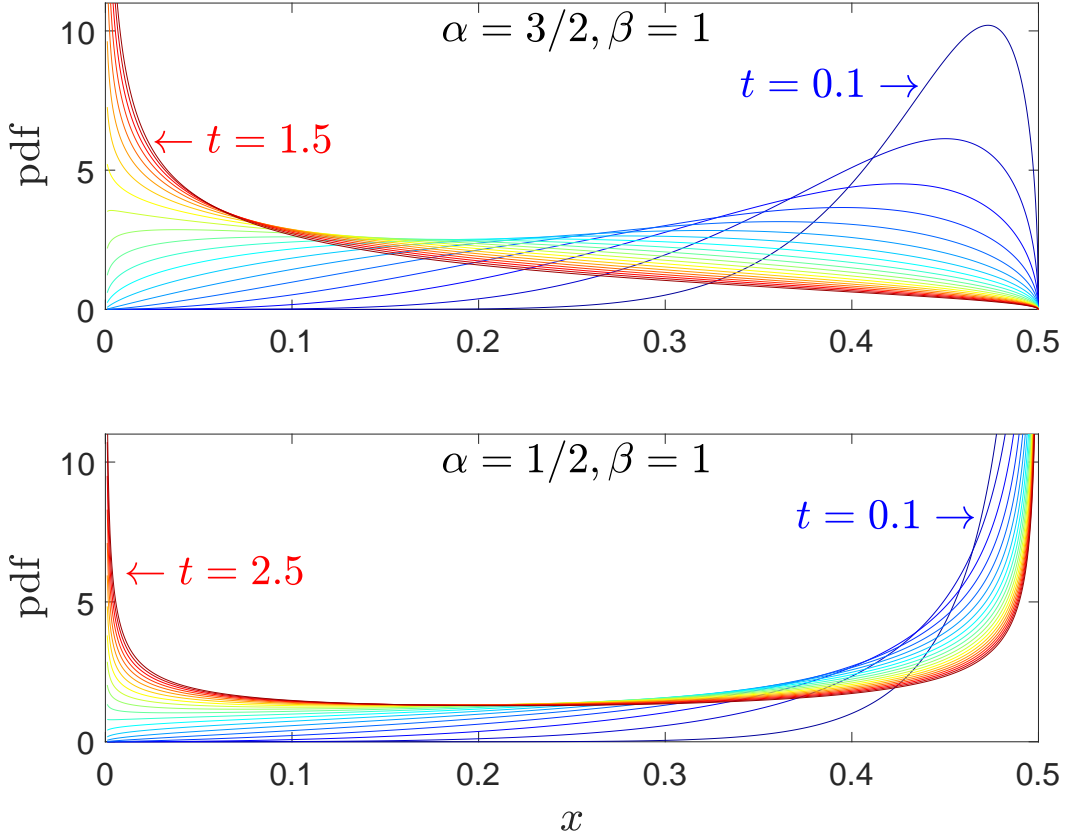}
\caption{Twenty PDFs of the covariance function $2^{-1}\exp(-\Lambda t), \Lambda
\deq\mathcal G(\alpha,\beta)$ for time $t$ changing linearly in the short memory
(top) and long memory (bottom) regimes.}
\label{f:pdfCov}
\end{figure}

The function \Ref{eq:rcOU} would be observed if we calculated the ensemble
average of $V_\Lambda(\tau)V_\Lambda(\tau+t)$ for some $\tau$. If instead the
covariance function would be estimated as a time average over individual
trajectories, the Birkhoff theorem determines that the result would be a
random variable, equal to the conditional covariance
\begin{equation}
\label{eq:taCov}
\overline{r_{V_\Lambda}(t)}\equiv\lim_{\mathcal T\to\infty}\f{1}{\mathcal T}\int_0^{\mathcal T}V_\Lambda(
\tau)V_\Lambda(\tau+t)\dd\tau= r_{V_\Lambda}(t|\Lambda)=\f{1}{2}\e^{-\Lambda t}.
\end{equation}
It is straightforward to calculate the PDF of this distribution,
\begin{equation}
\label{eq:expLbdpdf}
p_{\overline{r_{V_\Lambda}}}(x,t)=\f{2}{\Gamma(\alpha)}\lt(\beta/t\rt)^\alpha
|\ln(2x)|^{\alpha-1}(2x)^{\beta/t-1},\quad 0<x<1/2.
\end{equation}
The mean value of this quantity is given by result \Ref{eq:rcOU}. This PDF is zero
in the point $x=1/2$ if $\alpha>1$ but has a logarithmic singularity at $x=(1/2)^-$
if $\alpha<1$ (that is, in the long-memory case). It is zero in $x=0$ for $t<\beta$
as in expression \Ref{eq:expLbdpdf} any power law dominates any power of the
logarithm. For $t>\beta$ there is a singularity at $x=0^+$ which approaches the
asymptotics $x^{-1}|\ln x|^{\alpha-1}$ as $t\to\infty$. This behaviour can be
observed in Figure \ref{f:pdfCov}, illustrating how the probability mass moves
from $(1/2)^-$ to $0^+$ as time increases.

As we already know, the compound Ornstein-Uhlenbeck process is non-Gaussian. Let
us follow up on this property in more detail. To study the characteristic function
we need to calculate the average in equation \Ref{eq:phi}, which is actually the
moment generating function for the random variable $\exp(-\Lambda t)$. Some
approximations can be made. First, let us assume that $\Lambda t$ is small in the
sense that the probability that this variable is larger than some small $\epsilon>0$
is negligible. In this regime we can approximate $\exp(-\Lambda t)\approx1-\Lambda
t$ and find
\begin{align}
\label{eq:apprt0}
\nonumber
\phi_\Lambda(\bd\theta,t)&\approx \e^{-\f{1}{4}\theta_1^2}\e^{-\f{1}{4}\theta_2^2}
\E\lt[\e^{-\f{1}{2}\theta_1\theta_2(1-\Lambda t)}\rt]\\
&=\e^{-\f{1}{4}(\theta_1+\theta_2)^2}\E\lt[\e^{\f{1}{2}\theta_1\theta_2\Lambda
t}\rt]\nonumber\\
&=\e^{-\f{1}{4}(\theta_1+\theta_2)^2}\f{1}{(1-t\theta_1\theta_2/(2\beta))^{\alpha}},
\quad t\to 0^+.
\end{align}
The first factor describes a distribution of $V_\Lambda(\tau)=V_\Lambda(\tau+t)$.
So in our approximation we assume that the values in the process between short
time delays are nearly identical and the multiplicative correction $(1-t\theta_1
\theta_2/(2\beta))^{-\alpha}$ is non-Gaussian.

The second type of approximation can be made for long times $t$ when $\exp(-\Lambda
t)\approx 0$. In this case
\begin{align}
\nonumber
\phi_\Lambda(\bd\theta,t)&\approx\e^{-\f{1}{4}\theta_1^2}\e^{-\f{1}{4}\theta_2^2}
\E\lt[1-\f{1}{2}\theta_1\theta_2\e^{-\Lambda t}\rt]\\
&=\e^{-\f{1}{4}\theta_1^2}\e^{-\f{1}{4}\theta_2^2}\lt(1-\f{1}{2}\theta_1\theta_2
\E\lt[\e^{-\Lambda t}\rt]\rt)\nonumber\\
&=\e^{-\f{1}{4}\theta_1^2}\e^{-\f{1}{4}\theta_2^2}\lt(1-\f{\theta_1\theta_2}{2
(1-t/\beta)^\alpha}\rt),\quad t\to\infty.
\end{align}
Now we treat the values $V_\Lambda(\tau)$ and $V_\Lambda(\tau+t)$ as
nearly independent, the small correction is once again non-Gaussian. Apart from
the approximations, the exact formula for $\phi_\Lambda$ can be provided using
the series
\begin{equation}
\label{eq:phiSeries}
\E\lt[\e^{-\f{1}{2}\theta_1\theta_2\e^{-\Lambda t}}\rt]=\sum_{k=0}^\infty
\f{(-1)^k}{2^kk!}(\theta_1\theta_2)^k\E\lt[\e^{-k\Lambda t}\rt]=\sum_{k=0}
^\infty \f{(-1)^k}{2^kk!}(\theta_1\theta_2)^k\f{1}{(1+kt/\beta)^\alpha},
\end{equation}
which is absolutely convergent.

\begin{figure}
\centering
\includegraphics[width=16cm]{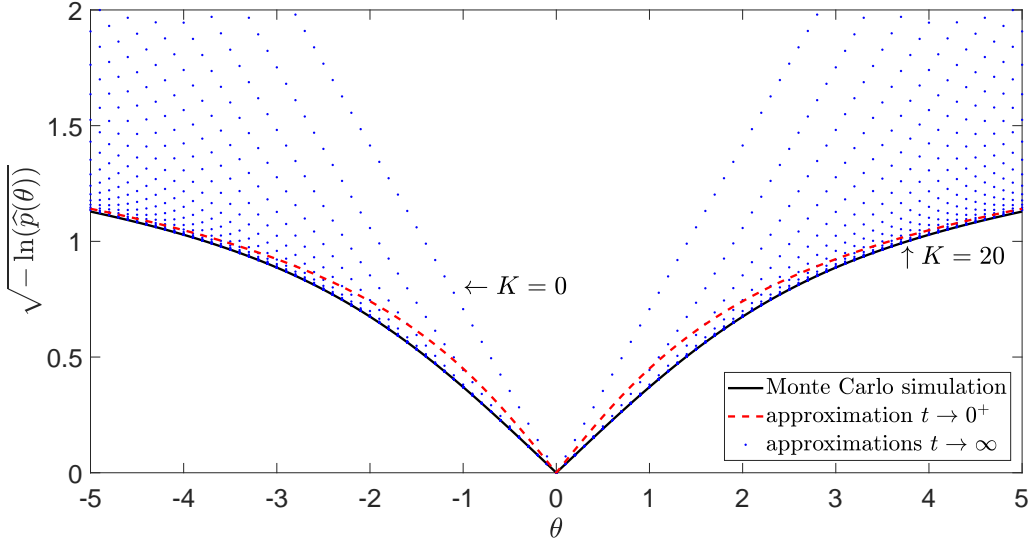}
\caption{Empirical characteristic function (solid black line) calculated from
Monte Carlo simulated for $\Delta V_\Lambda(\tau,1),\Lambda\deq\mathcal G(1/2,1)$;
sample size was $10^6$. The red dashed line represents approximation
\Ref{eq:apprt0}, the blue dotted lines are the approximations based on
equation \Ref{eq:apprtinfty} for $K=0,1,\ldots$, where $20$ terms in the Taylor
series were taken along.}
\label{f:fChar}
\end{figure}

Note that for the specific choice $\theta_1=\theta$, $\theta_2=-\theta$ the
function $\phi_\Lambda$ is a Fourier transform of the probability density of
the increment $\Delta V_\Lambda(\tau,t)\defeq V_\Lambda(\tau)-V_\Lambda(\tau+t)$,
which therefore equals
\begin{equation}
\label{eq:apprtinfty}
\widehat p_{\Delta V_\Lambda(\tau,t)}(\theta) = \e^{-\f{\theta^2}{2}}\sum_{k=0}^
\infty \f{\theta^{2k}}{2^k k!}\f{1}{(1+kt/\beta)^\alpha}.
\end{equation}
Clearly, any increment of $V_\Lambda$ is non-Gaussian. This is demonstrated
in Figure \ref{f:fChar}, where we show $\sqrt{-\ln(\widehat p_{\Delta V_\Lambda(
\tau,1)}(\theta))}$ on the y-axis. In this choice of scale Gaussian distributions
are represented by straight lines. The concave shape of the empirical estimator
calculated using Monte Carlo simulation shows that the process $V_\Lambda$ is
indeed non-Gaussian. In the same plot we present the two types of approximations
of $\widehat p_{\Delta V_\Lambda(\tau,1)}$: for $t\to 0^+$ we have equation
\Ref{eq:apprt0}, which reflects well the tails $\theta\to\pm\infty$, and
for $t\to\infty$ we see that with several terms of the series \Ref{eq:apprtinfty}
a good fit for $\theta\approx 0$ is obtained.
 
It may appear counter-intuitive that the values $V_\Lambda(t)$, which are all
exactly Gaussian, are sums of non-Gaussian variables. If the increments were
independent that would be impossible, here their non-ergodic dependence
structure allows for this unusual property to emerge. However, the they are still
conditionally Gaussian with variance
\begin{equation}
\E\lt[\Delta V_\Lambda(\tau,t)^2|\Lambda\rt]= \lt(1-\e^{-\Lambda t}\rt).
\end{equation}
The non-Gaussianity is prominent for short times $t$. As $t$ increases, the
distribution of $\Delta V_\Lambda(\tau,t)$ converges to a Gaussian with
unit variance.

The non-Gaussian memory structure of the velocity $V_\Lambda$ affects also
the distribution of the position $X_\Lambda$, which, using result
\Ref{eq:condmsd}, for large $t$ becomes
\begin{align}
\label{eq:Xpdf}
p_{X_\Lambda}(x,t)&\approx\f{1}{\sqrt{\pi t}}\E\lt[\sqrt{\Lambda}\e^{-x^2\Lambda/t}
\rt]=\f{\beta^\alpha}{\Gamma(\alpha)\sqrt{\pi t}}\int_0^\infty\lambda^{\alpha-1/2}
\e^{-(x^2/t+\beta)\lambda}\dd\lambda\nonumber\\
&=\f{\Gamma(\alpha+1/2)}{\sqrt{\pi}\Gamma(\alpha)}\f{(\beta t)^{\alpha}}{(x^2+\beta
t)^{\alpha+1/2}}.
\end{align}
For $\alpha=1/2$ and long $t$ the position process $X_\Lambda$ is approximately
Cauchy distributed. For short $t$ it is nearly Gaussian distributed with variance
$t^2/4$. In general the above formula is a PDF of the Student's T-distribution
type, although unusual in the sense that most often it arises in statistics where
it is parametrised only by positive integer values of $\alpha$. The parameter
$\alpha$ in the above expression determines the decay of the tails of the PDF,
which, as we can see, scale like $x^{-2\alpha-1}$: the parameter $\beta$ rescales
time, but in an inverse manner compared to its action on $V_\Lambda$.

It may therefore seem that for $\alpha\le 1/2$ the process $X_\Lambda$ may not
have a finite second moment, however, this is not true. In Section \ref{s:ssGLE}
we made the general remark that the MSD of the superstatistical GLE is necessarily
finite, see the comments below equation \Ref{eq:msdGen}. In our current case
$\delta^2_{X_\lambda}(t)$ is given in expression \Ref{eq:condmsd}, and thus
\begin{equation}
\lim_{\lambda\to 0^+}\delta^2_{X_\lambda}(t)=t^2/4,\quad  \lim_{\lambda\to
\infty}\delta^2_{X_\lambda}(t)=0.
\end{equation}
This is indeed a bounded function of control parameter $\lambda$, and the MSD
of $X_\Lambda$ must be finite for any distribution of $\Lambda$.  Moments of higher even order can be expressed as
\begin{equation} \label{eq:bound}
\E[X_\Lambda(t)^{2n}]=\prod_{k=2}^n(2k-1)\times\E\lt[(\delta^2_{X_\Lambda}(t|\Lambda)\big)^n\rt],
\end{equation}
so they are all also finite as well. Integrating
twice relation \Ref{eq:rcOU} for Gamma distributed $\Lambda$ it can be shown that
\begin{equation}
\delta^2_{X_\Lambda}(t)=\E[\delta^2_{X_\Lambda}(t|\Lambda)]=\f{\beta^2}{2}\f{(1+
t/\beta)^{2-\alpha}+(\alpha-2)t/\beta-1}{(1-\alpha)(2-\alpha)}.
\end{equation}
This describes superdiffusion for $0<\alpha<1$ and normal diffusion for $1\le
\alpha$ in agreement with the more general theory discussed below equation
\Ref{eq:condmsd}.

Similar, a somewhat longer calculation yields
\begin{align}
& \E\lt[\big(\delta^2_{X_\Lambda}(t|\Lambda)\big)^2\rt]=\f{\beta^4}{4}\f{1}{
(\alpha-4)(\alpha-3)(\alpha-2)(\alpha-1)}\nonumber\\
&\times\Big((\alpha-4) (\alpha-3) (t/\beta)^2-2 (\alpha-4) t/\beta+1\nonumber\\
& + 2 (\alpha-4) (t/\beta) (1+t/\beta)^{3-\alpha}-2
(1+t/\beta)^{4-\alpha}+(1+2 t/\beta)^{4-\alpha}\Big)
\end{align}
which determines the  asymptotics of the ergodicity breaking parameter \Ref{eq:EB}
\begin{equation}
\text{EB}(t)\sim\bg{cases}
\f{(1-\alpha)(2-\alpha)}{(4-\alpha)(3-\alpha)}\big(2
\alpha-10+2^{4-\alpha}\big)\big(\f{t}{\beta}\big)^{\alpha}, & \alpha<1\\
\f{(\alpha-1)}{(4-\alpha)(3-\alpha)(2-\alpha)}\big(10-2
\alpha-2^{4-\alpha}\big)\big(\f{t}{\beta}\big)^{2-\alpha}, & 1<\alpha<2\\
\f{1}{\alpha-2}, & 2<\alpha
\end{cases}
\end{equation}
at $t\to\infty$. Additionally, in this model it is easy to check that $3\times
(\text{EB}(t)+1)$  is the kurtosis of $X_\Lambda(t)$, that is $\E[X_\Lambda(t)^4]
/(\E[X_\Lambda(t)^2])^2$. This is one of the measures of the thickness of the
tails of a distribution which for any one-dimensional Gaussian distribution equals
3. Here, the distribution is clearly non-Gaussian, but it is hard to judge the
tail behaviour using the kurtosis. This is due to the fact that $p_{X_\Lambda}$ converges to a power-law, yet according to result \Ref{eq:bound} the tails of $p_{X_\Lambda}$ must decay faster than any power, symbolically $p_{X_\Lambda}(x,t)=\mathcal O(x^{-\infty})$ for any $t$. Therefore
the PDF's tails are always truncated, but is not noticeable observing moments, which are affected by the finite range in which PDF becomes close to the power law.

\begin{figure}
\centering
\includegraphics[width=16cm]{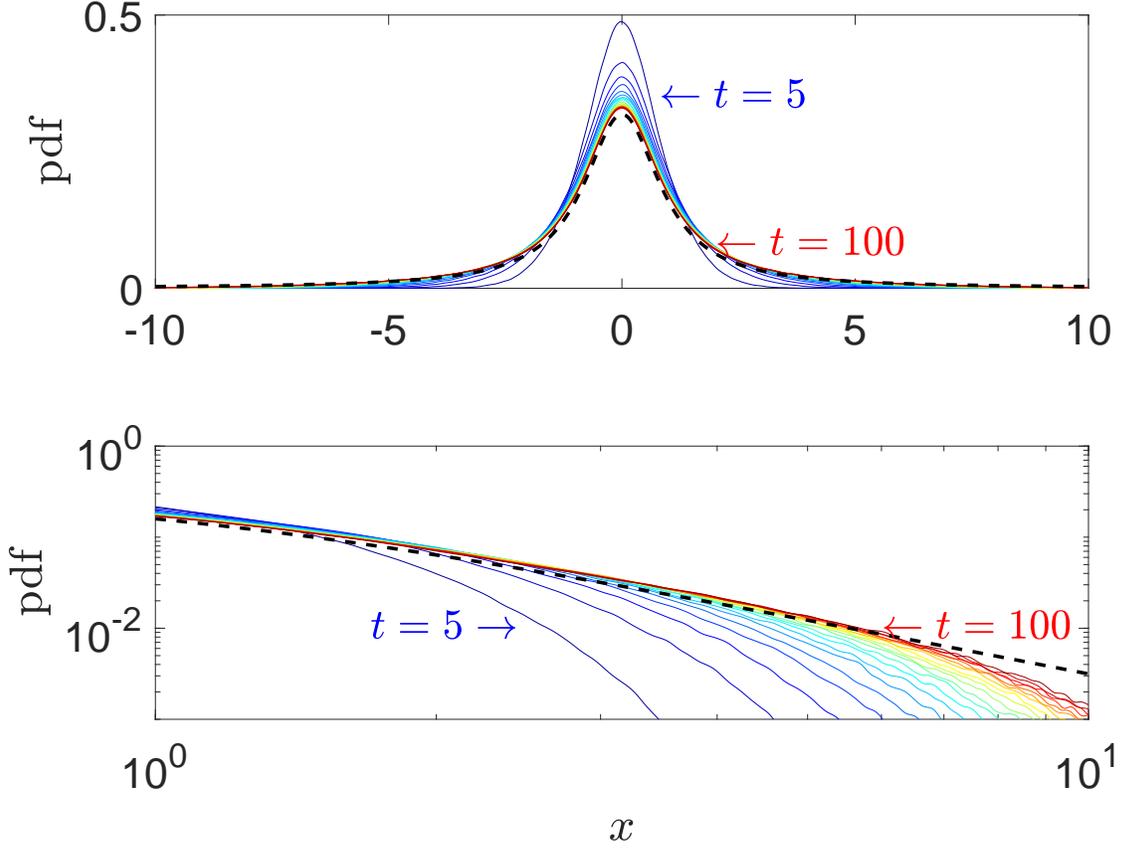}
\caption{Kernel density of variables $X_\Lambda(t)/\sqrt{t}$ estimated
for $t=5,10,\ldots,100$ (solid lines) and $\Lambda\deq\mathcal G(1/2,1)$
versus the Cauchy PDF (dashed line). Sample size was $10^6$. Both convergence to Cauchy distribution \Ref{eq:Xpdf} and the $\mathcal O(x^{-\infty})$ truncation of the tails \Ref{eq:bound} can be observed.}
\label{f:pdfX}
\end{figure}

The asymptotical properties of $X_\Lambda(t)$ are illustrated in Figure
\ref{f:pdfX}, where we show the PDFs of the rescaled position position
$X_\Lambda(t)/\sqrt{t}$ simulated with $\alpha=1/2,\beta=1$ and calculated
using kernel density estimator. In agreement with result \Ref{eq:Xpdf}, the
limiting distribution is of Cauchy type. At the same time for all finite $t$
the tails of the PDF remain truncated: as time increases this truncation is
moved more away into $x=\pm\infty$, and as a result the MSD increases as
$t^{2-\alpha}$.

This is an illustration of a more general rule: equation \Ref{eq:Xpdf} is a
Laplace transform of $\sqrt{\Lambda}$, so any $\Lambda$ with power law
$p_\Lambda(\lambda)\sim \lambda^{\alpha-1},\lambda\to 0^+$ will result in a
power law $x^{-2\alpha-1}$ as limiting distribution of $X_\Lambda$. But at the
same time the superstatistical Langevin equation preserves the finiteness of moments, which stems from the Hamiltonian derivation of the GLE. Therefore
this model reconciles power law tails of the observed distribution with a finite
second moment by naturally introducing truncation moving to $\pm\infty$
as $t\to\infty$.

\section{More complex memory types}
\label{morecomplex}

We here analyse the behaviour of the superstatistical GLE \Ref{eq:ssGLE}, which
is non-Markovian and may be used to model more complex types of memory structure.
We study the two important cases of exponential and power law shapes for the kernel.

\subsection{Exponential kernel GLE}

The covariance function of force in exponential kernel GLE has the conditional form
\begin{equation}
r_{\xi_{A,B}}(t|A,B)= B^2\e^{-2At},\quad A,B>0.
\end{equation}
This particular parametrisation is chosen for convenience, it will simplify the
formulas later. The stochastic force $\xi_{A,B}$ in this model is the compound
Ornstein-Uhlenbeck process considered in Section \Ref{s:cOU}, additionally
rescaled by random coefficient $B^2$. It may be represented in time space
or Fourier space as
\begin{equation}
\xi_{A,B}(t)= B^2\int_{-\infty}^t\xi(\tau)\e^{-2A(t-\tau)}\dd\tau=\int_{-\infty}
^\infty\widehat\xi(\omega)\f{B^2}{\I\omega+2A}\e^{\I\omega t}\dd\omega.
\end{equation}
We first solve the corresponding deterministic model. The Laplace transform
of the Green's function can be easily obtained in the form
\begin{equation}
\label{eq:GAB}
\widetilde G_{A,B}(s|A,B)=\f{1}{s+\f{B^2}{s+2A}}=\f{s+2A}{(s+A)^2-(A^2-B^2)}.
\end{equation}
Its Laplace inverse is a conditional covariance function, which is the sum of two exponential functions,
\begin{align}
\label{eq:rABFull}
r_{V_{A,B}}(t|A,B)&=\f{1}{2}\lt(1-\f{A}{\sqrt{A^2-B^2}}\rt)\e^{-(A+\sqrt{A^2
-B^2})t}\nonumber\\
&+\f{1}{2}\lt(1+\f{A}{\sqrt{A^2-B^2}}\rt)\e^{-(A-\sqrt{A^2-B^2})t}.
\end{align}
In the case $A=B$ a division by 0 appears, so the above formula should be
understood as a limit $A\to B$. In any case we can calculate the MSD using
relation \Ref{eq:msdGen} and obtain
\begin{align}
&\delta^2_{X_{A,B}}(t|A,B) = 4\f{A}{B^2} t \nonumber\\
&+\f{1}{\sqrt{A^2-B^2}}\e^{-At}\lt(\f{\sqrt{A^2-B^2}-A}{(A+\sqrt{A^2-B^2})^2}
\e^{\sqrt{A^2-B^2}t}+\f{\sqrt{A^2-B^2}+A}{(A-\sqrt{A^2-B^2})^2}\e^{-\sqrt{A^2
-B^2}t}\rt)\nonumber\\
&-\f{8A^2}{B^4}+2B^2.
\end{align}
As we can see the asymptotical behaviour of the MSD at $t=0^+$ and $t=\infty$
is very similar to that of the compound Ornstein-Uhlenbeck process. For
$\E[A/B^2]<\infty$ this GLE models normal diffusion with a random diffusion
coefficient. When this condition is not fulfilled it may model superdiffusion
determined by the power-law tails of the covariance function, compare
relation \Ref{eq:msdLapl} and the discussion below.

The behaviour of this system greatly depends on whether $A<B,A=B$ or $A>B$. All
ensemble averages can be separated between these three classes
\begin{equation}
\E[\bd\cdot]=\E[\bd\cdot|A<B]\pr(A<B)+\E[\bd\cdot|A=B]\pr(A=B)+\E[\bd\cdot|A>B]
\pr(A>B),
\end{equation}
so even if all three regimes can be present in a physical system, we can
model them separately and average the results at the end. We start the analysis
with the simplest case.

\subsubsection{Critical regime $A=B$.}

Taking the limit $A\to B$ in expression \Ref{eq:rABFull} or calculating the
inverse Laplace transform from equation \Ref{eq:GAB} with $A=B$ we determine
the form of the conditional covariance within this critical regime,
\begin{equation}
r_{V_{A}}(t|A)=(1+At)\e^{-A t}.
\end{equation}
The behaviour of the resulting solution $V_A$ is very similar to that of the
compound Ornstein-Uhlenbeck process. The differences are mostly technical. For
example, if $A=A_1+A_2$ for some independent $A_1$ and $A_2$, then
\begin{equation}
r_{V_{A_1+A_2}}(t)=r_{V_{A_1}}(t)r_{V_{A_2}}(t)-\E\lt[tA_1\e^{-A_1t}\rt]\E\lt
[tA_2\e^{-A_2t}\rt].
\end{equation}
Therefore, for instance if $A>a_0$ we can write $A=A'+a_0, A'>0$ and
the covariance function becomes truncated by $a_0t\exp(-a_0 t)$.

The formula for $r_{V_{A}}$ consist of two terms. Function $A t\exp(-A t)$
has a thicker tail, but the asymptotic behaviour of $r_{V_A}$ is determined by
the distribution of small values $A\approx 0$, so it is not clear which term
is most important in that regard. If we assume $p_A(a)\sim a^{\alpha-1}$ then
\begin{equation}
r_{V_{A}}(t)\sim \Gamma(\alpha)t^{-\alpha}+t\Gamma(\alpha+1)t^{-\alpha-1}=
(\alpha+1)\Gamma(\alpha)t^{-\alpha},
\end{equation}
so actually both terms have comparable influence over the resulting tails
of the covariance.

\subsubsection{Exponential decay regime $A>B$.}

In this case covariance function is a sum of two decaying exponentials. Because
of this $A>a_0$ results in an exponential truncation by $\exp(-a_0t)$ of the
associated covariance, but this time there is no simple rule to determine the
behaviour of this function for $A=A_1+A_2$. Instead let us analyse expression
\Ref{eq:rABFull} in more detail. The first exponential has a negative amplitude,
the second one a positive amplitude. In additional the second exponential always
has a heavier tail, as its exponent includes the difference of positive terms,
$A-\sqrt{A^2-B^2}$, whereas the other exponent includes a sum. Thus we expect
that the exponent with $A+\sqrt{A^2-B^2}$ cannot lead to a slower asymptotics
than the one containing $A-\sqrt{A^2-B^2}$.
	
Given this reasoning let us change the variables in the form
\begin{equation}
A' = A-\sqrt{A^2-B^2}=\f{B^2}{A+\sqrt{A^2-B^2}},\quad A=\f{B^2+A'^2}{2A'}.
\end{equation}
The new parameter $A'$ attains the value $A'=B$ for $A=B$ and decays monotonically to 0 as $A\to\infty$. Note that for small values of $A'$, $A'\approx B^2/(2A)$,
so the tail behaviour of $A$ determines the distribution of $A'$ at $0^+$; in
particular, a power law shape of the former is equivalent to a power law shape
of the latter. Using the parameters $A'$ and $B$ the covariance function can be
expressed as
\begin{equation}
r_{V_{A', B}}(t|A',B)=\f{B^2}{B^2-A'^2}\e^{-A' t}-\f{A'^2}{B^2-A'^2}
\e^{-\f{B^2}{A'}t}.
\end{equation}
As $A'<B$ the variables $A'$ and $B$ cannot be independent unless $B$ is
deterministic. In that latter case $B=b$ and for $A'$ concentrated around $0^+$
we have that
\begin{equation}
r_{V_{A',b}}(t)\sim\E\lt[\f{b^2}{b^2-A'^2}\e^{-A't}\rt]=\int_0^bp_{A'}(a)\f{b^2}{
b^2-a^2}\e^{-at}\dd a,
\end{equation}
so the asymptotical behaviour of this model is again in analogy to that of the
compound Ornstein-Uhlenbeck process. In particular, $p_{A'}(a)\sim a^{\alpha-1}$,
when $a\to 0^+$ (equivalently $p_A(a)\sim 2a^{-1-\alpha}$ for $a\to\infty$)
implies the emergence of a power law, $r_{V_{A',b}}(t)\sim \Gamma(\alpha)t^{
-\alpha}$. This asymptotic does not depend on the exact choice of $b$, which
means that the scale of the stochastic force does not affect the tails of the
memory and the influence of $b$ only matters at short times.

When both $A'$ and $B$ are random their dependence may potentially be quite
complex and influence the tails of the covariance in unpredictable ways. It
can only be studied under some simplifying assumptions. We want to require
some sort of independence between $A'$ and $B$ for small values of $A'$,
which determine the asymptotics of $\exp(-At)$. So, let us denote $B' \defeq
A'/B$, which is a random variable that must be less than 1, but may be supposed
to be independent from $A'$. Using variables $A'$ and $B'$ the covariance
function can be transformed into
\begin{equation}
r_{V_{A', B'}}(t|A',B')=\f{1}{1-B'^2}\e^{-A' t}-\f{B'^2}{1-B'^2}\e^{-\f{A'}{B'^2}t}.
\end{equation}
In this form the influence of $A'$ and $B'$ is mostly factorised, the only
remainder is $A'/B'^2$ in the second exponent. This leads to some immediate
consequences. If the PDF of $A'$ is supported on the interval $[a_1,a_2]$, and
$m\le p_{A'}(a)\le M$, then straightforward integration yields
\begin{equation}
\label{eq:expDecBoundedAs}
\E\lt[\f{1}{1-B'^2}\rt]\f{m}{t}\e^{-a_1 t}\lesssim r_{V_{A',B'}}(t)\lesssim
\E\lt[\f{1}{1-B'^2}\rt]\f{M}{t}\e^{-a_1 t}.
\end{equation}
Power law tails appear when $p_{A'}(a)\sim a^{\alpha-1},a\to 0^+$. The conditional
asymptotics then reads
\begin{equation}
r_{V_{A', B'}}(t|B')\sim\f{1}{1-B'^2}\Gamma(\alpha)t^{-\alpha}-\f{B'^2}{1-B'^2}
\Gamma(\alpha)\lt(\f{t}{B'^2}\rt)^{-\alpha},
\end{equation}
so for the unconditional covariance we have
\begin{equation}
\label{eq:expDecPowerLaw}
r_{V_{A', B'}}(t)\sim \Gamma(\alpha)\E\lt[\f{1-B'^{2\alpha+2}}{1-B'^2}\rt]
t^{-\alpha}.
\end{equation}
Both types of asymptotics are similar to the behaviour of the compound
Ornstein-Uhlenbeck process, only with different scaling. For the same reason
$r_{V_{A', B'}}(t)$ is truncated under the same conditions as before if it
is a sum of independent $A'_1$ and $A'_2$,
\begin{equation}
\label{eq:expDecmlpl}
r_{V_{A', B'}}(t)\sim \text{const}\times r_{V_{A'_1, B'}}(t)r_{V_{A'_2, B'}}(t),\quad t\to\infty,
\end{equation}
where the constant depends on the distribution of $A'$ and $B'$.

\subsubsection{Oscillatory decay regime $A<B$.}

When the square root $\sqrt{A^2-B^2}$ is imaginary we can express the covariance
function as
\begin{equation}
r_{V_{A,B}}(t|A,B)=\lt(\cos\big(\sqrt{B^2-A^2}t\big)+\f{A}{\sqrt{B^2-A^2}}
\sin\big(\sqrt{B^2-A^2}t\big)\rt)\e^{-At}.
\end{equation}
This represents a trigonometric oscillation truncated by the factor $\exp(-At)$.
 When calculating the unconditional covariance, this function acts as a integral kernel on the distribution of $A$ and $B$. The exponential factor acts similarly to the Laplace transform, but oscillations introduce Fourier-like behaviour of this transformation. It can be observed in the
solutions of the corresponding GLE, which we will show below.

Tauberian theorems can be applied for the bound of the covariance, given by
the inequality
\begin{equation}
|r_{V_{A,B}}(t|A,B)|\le \f{1}{\sqrt{1-\lt(\f{A}{B}\rt)^2}}\e^{-At},
\end{equation} 
which we prove in Proposition \ref{prp:oscAs} in the Appendix, together with
other asymptotic properties. As a general rule it can be said that solutions
of the GLE in this regime have a covariance which decays no slower than the
covariance of the compound Ornstein-Uhlenbeck process with the same distribution. For example, for independent $A$ and $B$, when the
PDF of $A$ is bounded and supported on some interval $[a_1,a_2]$,
\begin{equation}
\label{eq:oscAsInterval}
|r_{V_{A,B}}(t)|\lesssim \E\lt[\f{1}{\sqrt{1-\f{a^2_2}{B^2}}}\rt]\f{1}{t}
\e^{-a_1 t},
\end{equation}
and for a power law at $a_1^+$, that is $A=a_1+A'$ with $p_{A'}(a)\sim
a^{\alpha+1}$ at $a\to 0^+$, the covariance is bounded by
\begin{equation}
\label{eq:oscAsTrunc}
|r_{V_{A,B}}(t)|\lesssim \E\lt[\f{1}{\sqrt{1-\f{a^2_1}{B^2}}}\rt]\Gamma(\alpha)
t^{-\alpha}\e^{-a_1 t},\quad t\to \infty.
\end{equation}
The scaling constants depend on the distance between the distributions of $A$
and $B$: the closer they are the larger is the multiplicative factor. If
$A/B\approx 1-\epsilon$ it is roughly $\epsilon^{-1/2}$.

\begin{figure}
\centering
\includegraphics[width=16cm]{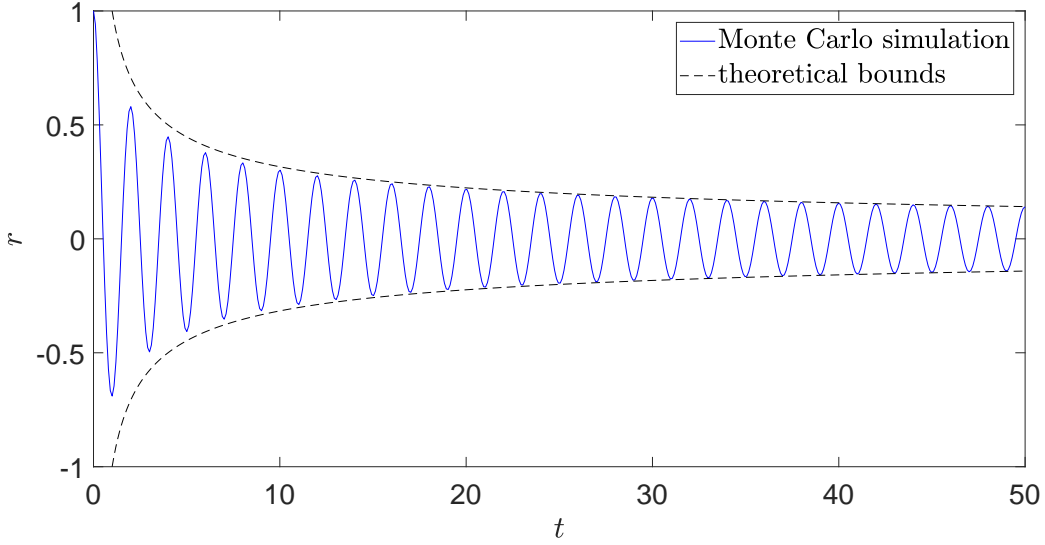}
\caption{Covariance function of the GLE (solid line) calculated using Monte Carlo
simulation, the sample size was $10^6$. Parameters are $B=\pi$, $A=\mathcal G
(1/2,1)$, the theoretical bounds are $\pm t^{-1/2}$ (dashed lines) as given in \Ref{eq:oscpl}.}
\label{f:oscCov}
\end{figure}

The question remains if this constraint is reached. The answer is yes, the
oscillations of $r_{V_{A,B}}(t|A,B)$ are asymptotically regular, that is, their
frequency becomes constant (exactly equal to $B$) at $t\to\infty$. Because of
this they are not influenced by averaging over $A$, so if $B$ is deterministic
$B=b$ and $A$ has power law $p_A(a)\sim a^{\alpha-1},a\to0^+$, we observe that
\begin{equation}
\label{eq:oscpl}
r_{V_{A,b}}(t)\sim \Gamma(\alpha)\cos(b t)t^{-\alpha}, t\to\infty.
\end{equation}
This behaviour can be seen in Figure \ref{f:oscCov} which demonstrates that
the convergence is relatively fast. During the Monte Carlo simulation the
parameter $B$ was fixed as $B=\pi$ and $A$ was taken from gamma distribution
$\mathcal G(1/2,1)$. For this distribution there exists $98.8\%$ chance that $A<\pi=B$ and the system is in the oscillatory regime, so it is indeed dominating the result, as shown in Figure \ref{f:oscCov}.

\subsection{Power law kernel GLE}

Our last example is the superstatistical GLE in which the force has a power law
covariance function, namely
\begin{equation}
r_{\xi_{H,Z}}(t|H,Z)=\f{Z}{\Gamma(2H-1)}t^{2H-2},\quad 0<H<1.
\end{equation}
The force process $\xi_{H,Z}$ is a fractional Brownian noise with random
index $H$, rescaled by the random coefficient $Z$. The factor $1/\Gamma(2H-1)$
is added for convenience and simplifies the formulas, however, its presence
does not change the outcome of our analysis.

The Green's function of this GLE is given by
\begin{equation}
\widetilde{G_{H,Z}}(s|H,Z)=\f{1}{s+Z s^{1-2H}}=\f{s^{2H-1}}{Z+s^{2H}}.
\end{equation}
In the last form we recognise the Laplace transform of a function from the
Mittag-Leffler class \cite{haubold}. The asymptotic of the conditional covariance
can be derived from Tauberian theorems or analysing the Mittag-Leffler function
directly \cite{haubold, gorenflo}
\begin{equation}
\label{eq:MLasympt}
r_{V_{H,Z}}(t|H,Z)=E_{2H}(-Zt^{2H})\sim \f{1}{Z\Gamma(1-2H)}t^{-2H},
\quad t\to\infty.
\end{equation}
From this formula we see that the distribution of $Z$ should not have an
influence on the covariance asymptotics. Further on we will assume that $Z$
is independent from $H$ and $\E[Z^{-1}]<\infty$. In the simple case when
$0<h_1<H<h_2$ and $H$ has a bounded PDF, that is $m\le p_{H}\le M$, one can
show that following bound holds
\begin{equation}
\label{eq:plInterval}
\f{m\E[Z^{-1}]}{2\Gamma(1-2h_1)}t^{-2h_1}\ln(t)^{-1}\lesssim r_{V_{H,Z}}(t)
\lesssim \f{M\E[Z^{-1}]}{2\Gamma(1-2h_1)}t^{-2h_1}\ln(t)^{-1},\quad t\to\infty.
\end{equation}
Therefore $r_{V_{H,Z}}\asymp t^{-2h_1}\ln(t)^{-1}$. The proof is given in
Proposition \ref{prp:MLas} i) in the Appendix. As usual, when $H$ has uniform
distribution on $[h_1,h_2]$ the asymptotics is stronger, $r_{V_{H,Z}}\sim
\E[Z^{-1}](2\Gamma(1-2h_1)(h_2-h_1))^{-1} t^{-2h_1}\ln(t)^{-1}$.

A more interesting situation occurs when $H$ is distributed according to
a power law. Noting that $t^{-2H}=\e^{-2H\ln(t)}$ one may suspect that
the resulting covariance would exhibit power-log tails. This intuition is
true. We will analyse case the when the index is $H=H'+h_0, h_0> 0, p_{H'}(h)\sim
h^{\alpha-1}$. By imposing $h_0>0$ we prohibit a situation when values of $H$
are arbitrarily close to $0^+$ because the Mittag-Leffler function diverges in this
limit, otherwise it is a continuous function of $H$. This problem corresponds
to the fact that for small $H$ the trajectories of $\xi_{H,Z}$ become very
irregular and as $H\to 0^+$ the solution of GLE is not well-defined. We
show in the Appendix that under these assumptions the asymptotics indeed has
power-log factor
\begin{equation}
\label{eq:powerLawAs}
r_{V_{H,Z}}(t)\sim\f{\E[Z^{-1}]\Gamma(\alpha)}{2^\alpha\Gamma(1-2h_0)}
t^{-2h_0}\ln(t)^{-\alpha}.
\end{equation}
Because we can take $h_0$ arbitrarily close to $0^+$ in this model we can
obtain tails which are very close to pure power-log shape.

To finish this section let us also comment on the properties of the position
process. Equation \Ref{eq:msdGen} describes the MSD as a second derivative
of the Green's function, so using its simple form in the Laplace space
\Ref{eq:msdLapl}
\begin{equation}
\widetilde{\delta^{2}_{X_{H,Z}}}(s|H,Z)=\f{s^{2H-3}}{Z+s^{2H}}.
\end{equation}
The inverse transform can be found using tables of two-parameter Mittag-Leffler
function, which also determines its asymptotics \cite{haubold, gorenflo}
\begin{equation}
\delta^{2}_{X_{H,Z}}(t|H,Z)=t^2E_{2H,3}(-Zt^{2H})\sim\f{1}{Z\Gamma(3-2H)}
t^{2-2H},\quad t\to\infty.
\end{equation}
The presence of the factor $1/Z$ means that this superstatistical GLE can model
anomalous diffusion with non-Gaussian PDF's. This dependence on $Z$ is the
same as for the parameter $\Lambda$ of the compound Ornstein-Uhlenbeck process,
so both exponential and power law tails can be present in this model in an
analogous way.

\begin{figure}
\centering
\includegraphics[width=16cm]{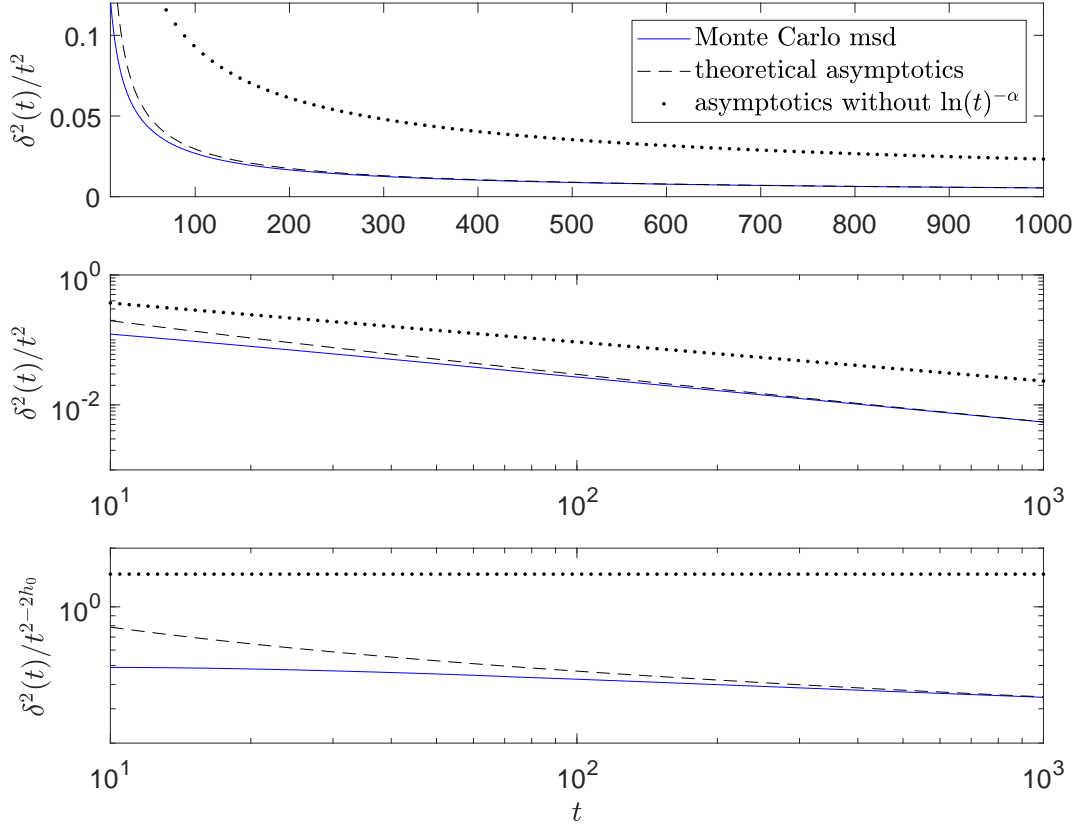}
\caption{MSD from Monte Carlo simulations (solid blue lines) for power law GLE
with fixed $Z=1$ and $0.3<H<0.5$ which has the power law form $p_{H}(0.3+h)\sim
\mathrm{const}\times h^{\alpha-1}$ with $\alpha=3/4$. The sample size was $10^3$.
The result is show on different scales, together with the asymptotic
\Ref{eq:powerLawAs} (dashed lines) and the same asymptotics without the factor
$\ln(t)^{-3/4}$ (dotted lines).}
\label{f:powerLawMSD}
\end{figure}

As for the asymptotic of $\delta^{2}_{X_{H,Z}}(t)$, the identical argument
as for the covariance can be used, so in this model the MSD of the form
\begin{equation}
\label{eq:msdAs}
\delta^{2}_{X_{H,Z}}(t)\sim\f{\E[Z^{-1}]\Gamma(\alpha)}{2^\alpha\Gamma(3-2h_0)}
t^{2-2h_0}\ln(t)^{-\alpha}
\end{equation}
is present for $0< h_0<1$ and $0<\alpha\le 1$. A numerical evaluation of
this behaviour is shown in Figure \ref{f:powerLawMSD} where we have taken the
subdiffusive case $H=3/10+H'$, with $p_{H'}(h)=\alpha5^{-\alpha} h^{\alpha-1}$
and $\alpha=3/4$, $0<H'<1/5$. The factor $t^2$ in expression \Ref{eq:msdAs}
does not depend on the particular form of the dynamics, so we divided all shown
functions by this factor to highlight the influence of $H$. As we can see the
convergence to the asymptotic behaviour is much slower than in the previous
examples, which stems from the fact that the Mittag-Leffler function converges
slowly to the power law
\begin{equation}
E_{2H,3}(-Zt^{2H})=\f{t^{-2H}}{Z\Gamma(3-2 H)}+\mathcal O(t^{-4H}).
\end{equation}
The inclusion of the log power law is significant, but may be difficult
to determine on the log log scale. It is demonstrated in the two lower panels
in Figure \ref{f:powerLawMSD}. The asymptotic MSD is concave on log log
scale, but the effect is not very prominent, and for the MSD estimated from
Monte Carlo simulations can be detected only on very long time scales. The
difference from  a power law is more visible if the lines are shown without
the factor $t^{-2h_0}$ (bottom panel), but $h_0$ may not be easy to estimate
for real systems. This comparison shows that different possible forms of decay
can be easily mistaken, so one should exert caution when analysing data suspected
to stem from such systems.

\section{Conclusions}
\label{summary}

We here studied the properties of the solutions of the superstatistical
generalised Langevin equation. This is based on the Gaussian GLE, but
includes random parameters determining the stochastic force. This
new type of processes has a number of properties that are unusual among
the models of diffusion. Firstly, the velocity is a stationary but
non-ergodic process. The behaviour of its time-averages can be studied,
for instance, using the time averaged covariance function \Ref{eq:taCov},
see also Figure \ref{f:pdfCov}. The resulting process is moreover not
Gaussian. At every point of time it has a Gaussian distribution but exhibits
a non-Gaussian structure of the memory, see equations \Ref{eq:apprt0} and
\Ref{eq:phiSeries}. One consequence of this is that it has non-Gaussian
increments \Ref{eq:apprtinfty}, also shown in Figure \ref{f:fChar}.

Secondly,
the position process at small times has a PDF which is approximately Gaussian,
but as time increases it converges to non-Gaussian PDF, as demonstrated in
Figure \ref{f:pdfX}. This limit distribution can exhibit commonly observed
exponential tails but also power-law tails \Ref{eq:Xpdf}. Even when the
limiting distribution does not have a finite second moment, at any given
finite time $t$ the position process does have a finite MSD. The observed
PDFs are always truncated. For short memory GLE, the MSD of the position
process is normal or superdiffusive, see \Ref{eq:condmsd} and the discussion
below. For power law memory models, anomalous diffusion with additional
log-power law may be observed \Ref{eq:msdAs}.

Various kinds of the GLE, distributions of shape parameters and the resulting
asymptotic properties of the covariance function $r_V$ based on the model
developed herein, are shown in Table \ref{t:covBeh}. The notation is chosen
as follows: $f(t)\sim g(t)$ when $f(t)/g(t)\to 1$, $f(t)\lesssim g(t)$ when
$f(t)\le h(t)\sim g(t)$, and $f(t)\asymp g(t)$ when $c_1 g(t)\lesssim
f(t) \lesssim c_2 g(t)$ for some constants $0<c_1<c_2$. The middle and left
columns can be slightly generalised if the distribution of shape parameters
contains a slowly varying factor, see Proposition \ref{prp:slowVar}.

\begin{table}
\centerline{\begin{tabular}{|c|c|c|c|}
 \cline{2-4}
\multicolumn{1}{c}{}  & \multicolumn{3}{|c|}{Shape parameter distribution} \\
\hline
$r_\xi$ & bounded on interval & power law at $0^+$ & power law + $c$, $c<1$\\
 \hline
 $\Lambda\delta(t)$   & $\asymp$ exponent $\times\ \f{1}{t}$  \Ref{eq:cOUBoundedAs} &$\sim$ pow. law \Ref{eq:cOUPowerLaw}&$\sim$ pow. law $\times$ exponent \Ref{eq:cOUmultipl}\\
 $\e^{-\f{1+A'^2}{A'} t}$ &$\asymp$  exponent $\times\ \f{1}{t}$ \Ref{eq:expDecBoundedAs}    &$\sim$ pow. law \Ref{eq:expDecPowerLaw} &$\sim$ pow. law $\times$ exponent \Ref{eq:expDecmlpl} \\
$\e^{-2A t}$ &  $\lesssim$  exponent  $\times\ \f{1}{t}$ \Ref{eq:oscAsInterval} &$\sim$ osc. pow. law \Ref{eq:oscpl}&$\lesssim$ pow. law $\times$ exponent \Ref{eq:oscAsTrunc}\\
 $t^{2H-2}$ &$\asymp$ pow. law $\times\ \f{1}{\ln(t)}$  \Ref{eq:plInterval}& ill-defined&$\sim$ pow. law $\times$  pow.-log law \Ref{eq:powerLawAs}\\
 \hline
\end{tabular}}
\caption{Different asymptotics of the covariance function $r_V$ for different
GLE: memoryless, exponential kernel in decay regime (under convenient
parametrisation), exponential kernel in oscillating exponential decay regime,
and power law. Different distributions of the shape parameters $\Lambda,A',A,H$
are considered}
\label{t:covBeh}
\end{table}

We see our work as a part of the development of the superstatical approach
to the modelling of diffusive processes in complex media. The assumption that
the stochastic force driving the GLE can change its properties from one
localised trajectory to another appears quite natural. Moreover, the
superstatistical GLE can simultaneously explain the presence of non-Gaussian
distributions and normal or anomalous type of diffusion. The properties of the
solutions listed above are specific enough to clearly distinguish this model
from possible alternatives, in particular, from the presence of non-Gaussian
PDFs caused by random rescaling of the process, which does not change the type
of memory that is observed in the system. We show how in our model the ensembles
of short memory trajectories can, concertedly, give rise to a long memory. This
dependence is highly non-Gaussian despite the Gaussian PDF of the velocity process.
The presence of such peculiar phenomena in a physical model is an interesting
theoretical finding in its own right. Its main relevance, however, lies in
the description of stochastic data, highlighting the importance of comprehensive
Gaussianity checks in data analysis.

We also note that the classical Langevin equation, considered in Section
\ref{s:cOU}, exhibits most of the properties specific for superstatistical
GLE---even the presence of power-log tails. This model has a very simple form
and at the same time allows for the derivation of not only asymptotic results,
but also many exact ones, given distribution of the dumping coefficient
or viscosity. Its most significant limitation is that it cannot model
subdiffusion. As such it should be considered a simple yet robust model for
data with linear or superdiffusive msd, power law or log-power law covariance
function and non-Gaussian PDFs.

Finally, let us add a caveat. Of course, any locally diffusing particle will
eventually reach the border of its domain characterised by a specific set of
diffusion parameters. For time ranges much longer than this dissociation
time, first a coarse-graining of the diffusion parameters will arise, and
eventually the diffusion will be governed by a Gaussian diffusion process
with a single, effective diffusion coefficient, similar to the observations
in the diffusing diffusivity model \cite{diffDiffChechkin}.

\ack

R M acknowledges financial support from DFG within project ME 1535/6-1. J {\' S}  and M M  acknowledge financial support from NCN HARMONIA 8 grant no. UMO-2016/22/M/ST1/00233.

\section{Appendix: Proofs}

We here provide several propositions with proofs, needed in the development
of the theory in the main body of the paper.

\bg{prp}\label{prp:cov}
The covariance function of any stationary solution $V$ of the GLE \Ref{eq:GLE} equals
\begin{equation}
r_V(t)=G(t), \quad t\ge 0,
\end{equation}
where $G$ is the corresponding Green's function of the GLE, and the MSD
of the position process $X(t)=\int_0^t\dd\tau\ V(t)$ is
\begin{equation}
\delta_X^2(t) = 2\int_0^t\dd\tau_1\int_0^{\tau_1}\!\dd\tau_2\ G(\tau_2).
\end{equation}
\end{prp}
\bg{proof} We will assume that the Green's function $G$ and kernel $K$ are
functions on $\R$ for which $G(t)=0$ and $K(t)=0$ for $t<0$ (such functions
are called ''casual''). Using this notation all integrals are defined on $\R$,
which simplifies calculations.

The solution of the GLE is stationary and has zero mean, so we can take $t>0$ and
calculate the covariance function as
\begin{align}
r_\xi(t)&=\E[V(0)V(t)]=\E\lt[\int_\R \dd\tau_1\ \xi(\tau_1)\ G(-\tau_1)\int_\R\dd\tau_2\ \xi(\tau_2) G(t-\tau_2)\rt]\nonumber\\
&= \int_\R\dd\tau_1\int_\R\dd\tau_2\ r_\xi(\tau_2-\tau_1) G(-\tau_1)G(t-\tau_2).
\end{align}
The covariance function $r_\xi$ is not casual, but it is symmetric, so it can
be represented as $r_\xi(\tau)=K(\tau)+K(-\tau)$. This formula fails only at
$\tau=0$, but it does not affect result of integration. The integral separates
into two parts, the first being
\begin{align}
I_1&=\int_\R\dd\tau_1\int_\R\dd\tau_2\ K(\tau_2-\tau_1) G(-\tau_1)G(t-\tau_2)
\nonumber\\
&=\int_\R\dd\tau_1\ G(-\tau_1)\int_\R\dd\tau_2'\ K(t-\tau_1-\tau_2')G(\tau_2')
\nonumber\\
&=-\int_\R G(-\tau_1)\dot G(t-\tau_1)\dd\tau_1,
\end{align}
where we used relation \Ref{eq:defG} in the interior of support of $G$. Similarly,
for the second integral,
\begin{align}
I_2&=\int_\R\dd\tau_1\int_\R\dd\tau_2\ K(\tau_1-\tau_2) G(-\tau_1)G(t-\tau_2)
\nonumber\\
&=\int_\R\dd\tau_2G(t-\tau_2)\int_\R\dd\tau_1'\ K(-\tau_2-\tau_1') G(\tau_1')
\nonumber\\
&=-\int_\R G(t-\tau_2)\dot G(-\tau_2)\dd\tau_2.
\end{align}
In $I_1$ and $I_2$ we can substitute $-\tau_1=\tau$ and $-\tau_2=\tau$. In their
sum we recognise the formula for integration by parts, which yields
\begin{equation}
r_V(t) = -G(t+\tau) G(\tau)\Big|_{\tau=0^+}^{\tau=\infty}= k_B T G(t) G(0^+)=G(t).
\end{equation}
Now, for the position process we find
\begin{equation}
\delta_X^2(t)=\int_0^t\dd\tau_1\int_0^t\dd\tau_2 \ r_V(\tau_2-\tau_1)=\int_0^t
\dd\tau_1\int_0^t\dd\tau_2 \ \big(G(\tau_2-\tau_1)+G(\tau_1-\tau_2)\big).
\end{equation}
Because of the symmetry between $\tau_1$ and $\tau_2$ the integral is twice the
term with $G(\tau_1-\tau_2)$, after substitution $\tau_1-\tau_2=\tau_2'>0$ we get
\begin{equation}
\delta_X^2(t) =2 \int_0^t\dd\tau_1\int_0^t\dd\tau_2 \ G(\tau_1-\tau_2) = 2\int_0^t\dd\tau_1\int_0^{\tau_1}\!\dd\tau_2' \ G(\tau_2').
\end{equation}
\end{proof}

\bg{prp}\label{prp:slowVar}

If $L$ is a slowly varying function at $0^+$, that is
\begin{equation}
\lim_{\lambda\to 0^+}\f{L(\lambda x)}{L(\lambda)}=1\quad \text{  for any } x>0,
\end{equation}
then
\bg{itemize}
\item[i)] A distribution of the form
\begin{equation}
p_\Lambda(\lambda)\sim \lambda^{\alpha-1} L(\lambda),\,\,\,\lambda\to0^+
\end{equation}
implies that the mean value of the exponential satisfies
\begin{equation}
\E\lt[\e^{-\Lambda t}\rt]\sim \Gamma(\alpha) t^{-\alpha}L(t^{-1}),\quad t\to\infty.
\end{equation}
\item[ii)] A distribution of $H$ of the form $H=h_1+H'$ with $h_1>0$
\begin{equation} 
p_{H'}(h)\sim h^{\alpha-1} L(h),\quad h\to0^+
\end{equation}
implies that the mean value of a power law satisfies
\begin{equation}
\E\lt[t^{-2H}\rt]\sim \f{\Gamma(\alpha)}{2^\alpha}t^{-2h_1}L(\ln(t)^{-1})
\ln(t)^{-\alpha},\quad t\to\infty.
\end{equation}
\end{itemize}
\end{prp}
\bg{proof}
We will only show ii), as the proof of i) is similar and simpler. We write
the integral for $\E[t^{-2(h_1+H')}]$, reformulate it as a Laplace transform
using $t^{-2H'}=\e^{-2H'\ln(t)}$, change variables and calculate the limit
\begin{align}
&\f{\E\lt[t^{-2(h_1+H')}\rt]}{t^{-2h_1}L(\ln(t)^{-1})\ln(t)^{-\alpha}}=\f{1}{
t^{-2h_1}L(\ln(t)^{-1})\ln(t)^{-\alpha}}\int_0^{1-h_1}p_{H'}(h)t^{-2(h_1+h)}\\d h
\nonumber\\
&=\int_0^{1-h_1}\f{p_{H'}(h)}{L(\ln(t)^{-1})\ln(t)^{-\alpha}}e^{-2h\ln(t)}\dd h=
\int_0^{\ln(t)(1-h_1)}\f{p_{H'}\lt(\f{h}{\ln t}\rt)}{L(\ln(t)^{-1})\ln(t)^{1-
\alpha}}e^{-2h}\dd h
\nonumber\\
&\xrightarrow{t\to\infty}\int_0^{\infty}h^{\alpha-1}e^{-2h}\dd h=\f{\Gamma(
\alpha)}{2^\alpha}.
\end{align}

\end{proof}

\bg{prp}\label{prp:oscAs}
Let $r_{V_{A,B}}$ be a covariance function corresponding to the solution of
a GLE with an exponential kernel in the oscillatory regime,
\begin{equation}
r_{V_{A,B}}(t|A,B)= \lt(\cos\big(\sqrt{B^2-A^2}t\big)+\f{A}{\sqrt{B^2-A^2}}
\sin\big(\sqrt{B^2-A^2}t\big)\rt)\e^{-At}.
\end{equation}
Then the following asymptotical properties hold:
\bg{itemize}
\item[i)] For $A$ with bounded PDF $p_A\le M$ supported on the interval $[a_1,
a_2]$, $a_2<B$ and independent of $B$, there exists the asymptotic bound
\begin{equation}
|r_{V_{A,B}}(t)|\lesssim \E\lt[\f{1}{\sqrt{1-\f{a_2^2}{B^2}}}\rt]\f{M}{t}
\e^{-a_1 t},\quad t\to\infty.
\end{equation}
\item[ii)] If additionally $A$ exhibits a power law behaviour at $a_1^+$, that
is, $A=a_1+A', p_{A'}(a)\sim a^{\alpha+1}$ for $a\to 0^+$, the asymptotic bound
can be refined to
\begin{equation}
|r_{V_{A,B}}(t)|\lesssim\E\lt[\f{1}{\sqrt{1-\f{a_1^2}{B^2}}}\rt]M\Gamma(\alpha)
t^{-\alpha}\e^{-a_1t}.
\end{equation}
\item[iii)]For $p_{A}(a)\sim a^{\alpha+1}$ at $a\to 0^+$ and deterministic $B=b$
the asymptotic limit of covariance function is
\begin{equation}
r_{V_{A,b}}(t)\sim \Gamma(\alpha)\cos(bt)t^{-\alpha},\,\,\, t\to\infty,
\end{equation}
which holds for all sequences of $t_k\to\infty$ which do not target zeros of
$\cos(bt)$, that is $|bt_k-l\pi+\pi/2|>\epsilon$ for all $k,l\in\N$ and some
$\epsilon>0$.
\end{itemize}
\end{prp}

\bg{proof}
We start from the simple inequality
\begin{align}
|r_{V_{A,B}}(t|A,B)|&=\lt|\sqrt{1-\lt(\f{A}{B}\rt)^2}\cos\big(\sqrt{B^2-A^2}t
\big)+\f{A}{B}\sin\big(\sqrt{B^2-A^2}t\big)\rt|\nonumber\\
&\times\f{1}{\sqrt{1-\lt(\f{A}{B}\rt)^2}}\e^{-At}\nonumber\\
&=\lt|\cos\lt(\sqrt{B^2-A^2}t-\arcsin\lt(\f{A}{B}\rt)\rt)\rt|\f{1}{\sqrt{1-
\lt(\f{A}{B}\rt)^2}}\e^{-At}\nonumber\\
&\le \f{1}{\sqrt{1-\lt(\f{A}{B}\rt)^2}}\e^{-At}.
\end{align}
This allows us to prove i), namely:
\begin{equation}
|r_{V_{A,B}}(t|B)|\le M\int_{a_1}^{a_2}\f{1}{\sqrt{1-\f{a^2}{B^2}}}\e^{-at}\dd a
\le\f{M}{\sqrt{1-\f{a_2^2}{B^2}}}\int_{a_1}^{\infty}\e^{-at}\dd a
=\f{M}{\sqrt{1-\f{a_2^2}{B^2}}}\f{1}{t}\e^{-a_1 t}.
\end{equation}
Averaging over $B$ yields the result. For $B$ with a distribution concentrated at
$a_2^+$ it may happen that
\begin{equation}
\E\lt[\f{1}{\sqrt{1-\f{a_2^2}{B^2}}}\rt]=\infty
\end{equation} 
and in this case point i) is a trivial statement. However, it is sufficient that
$B>a_2+\epsilon,\epsilon>0$ for this average to be finite and $\le 2/\epsilon$.

Proof of point ii) is similar,
\begin{equation}
|r_{V_{A,B}}(t|B)|\le M\int_0^\infty\f{p_{A'}(a)}{\sqrt{1-\f{(a+a_1)^2}{B^2}}}
\e^{-(a+a_1)t}\dd a\sim\f{M\Gamma(\alpha)}{\sqrt{1-\f{a_1^2}{B^2}}}t^{-\alpha}
\e^{-a_1t}.
\end{equation}

Proving iii) requires a more delicate reasoning. We write $r_{V_{A,b}}(t)$
as an integral and change variables $at\to a$, so that
\begin{align}
r_{V_{A,b}}(t)&=\int_0^\infty p_A(a)\lt(\cos\lt(\sqrt{b^2-a^2} t\rt)+\f{a}{
\sqrt{b^2-a^2}}\sin\lt(\sqrt{b^2-a^2}t\rt)\rt)\e^{-at}\dd a\nonumber\\
&=\f{1}{t}\int_0^\infty p_A\lt(\f{a}{t}\rt)\lt(\cos\lt(\sqrt{b^2t^2-a^2}\rt)
+\f{a}{\sqrt{b^2t^2-a^2}}\sin\lt(\sqrt{b^2t^2-a^2}\rt)\rt)\e^{-a}\dd a.
\end{align}
After change of variables the Fourier oscillations depend on the variable
$\sqrt{b^2t^2-a^2}$. In the limit $t\to\infty$ they converge to oscillations
with frequency $b$,
\begin{equation}
\big|\sqrt{b^2t^2-a^2}-bt\big|=\f{a^2}{\sqrt{b^2t^2-a^2}+bt}\xrightarrow{t\to
\infty} 0.
\end{equation}
It is crucial that this frequency does not depend on $a$. The cosine function
also converges to a cosine with frequency $b$,
\begin{align}
\f{\cos(\sqrt{b^2t^2-a^2})}{\cos(bt)}&=\f{\cos(\sqrt{b^2t^2-a^2}-b t+bt)}{\cos(b
t)}\nonumber\\
&=\f{\cos(\sqrt{b^2t^2-a^2}-b t)\cos(bt)+\sin(\sqrt{b^2t^2-a^2}-b t)\sin(bt)}{
\cos(bt)}\nonumber\\
&\xrightarrow{t\to\infty}\cos(0)\times1+\sin(0)\times\tan(bt)=1,\quad|\tan(bt)|
<\f{1}{\epsilon}.
\end{align}
Substituting this result into the integral for $r_{V_{A,b}}$ we obtain the
asymptotic
\begin{align}
\f{r_{V_{A,b}}(t)}{\cos(bt) t^{-\alpha}}&=\int_0^\infty p_A\lt(\f{a}{t}\rt)
t^{\alpha-1}\f{\cos(\sqrt{b^2t^2-a^2})+\f{a}{\sqrt{b^2t^2-a^2}}\sin(\sqrt{b^2
t^2-a^2})}{\cos(bt)}\e^{-a}\dd a\nonumber\\
&\xrightarrow{t\to\infty} \int_0^\infty a^{\alpha-1}\e^{-a}=\Gamma(\alpha)\dd a.
\end{align}
\end{proof}

\bg{prp}\label{prp:MLas}
Let $H$ and $Z$ be independent, $\E[Z^{-1}]<\infty$ and $\beta\ge 1$. Then the
following asymptotic properties of $\E[E_{2H,\beta}(-Zt^{2H})]$ hold
\bg{itemize}
\item[i)] If $H$ is supported on $[h_1,h_2]$ with $0<h_1<h_2\le 1$ and its PDF
is bounded, $m\le p_H(h)\le M$, then
\begin{equation}
\f{m\E[Z^{-1}]}{2\Gamma(\beta-2h_1)}t^{-2h_1}\ln(t)^{-1}\lesssim \E[E_{2H,\beta}
(-Zt^{2H})] \lesssim \f{M\E[Z^{-1}]}{2\Gamma(\beta-2h_1)}t^{-2h_1}\ln(t)^{-1}.
\end{equation}
\item[ii)] If additionally $H$ exhibits a power law behaviour at $h_1^+$, that
is, $H=h_1+H'$, $p_{H'}(h)\sim h^{\alpha+1}$ with $h\to 0^+$, a much stronger
asymptotic property holds,
\begin{equation}
\E[E_{2H,\beta}(-Zt^{2H})]\sim\f{\E[Z^{-1}]\Gamma(\alpha)}{2^\alpha\Gamma(\beta
-2h_1)}t^{-2h_1}\ln(t)^{-\alpha},\quad t\to \infty.
\end{equation}
\end{itemize}
\end{prp}
\bg{proof}
Because
\begin{equation}
\label{eq:MLasympt2}
E_{2H,\beta}(-Zt^{2H})\sim \f{1}{Z\Gamma(\beta-2H)}t^{-2H}
\end{equation}
the left hand side is a function witch has constant sign. For some small
$\epsilon>0$ and large enough $t$ we observe the inequality
\begin{equation}
\f{m(1-\epsilon)}{Z}\int_{h_1}^{h_2}\f{t^{-2h}}{\Gamma(\beta-2h)}\dd h
\le \E[E_{2H,\beta}(-Zt^{2H})|Z]\le\f{M(1+\epsilon)}{Z}\int_{h_1}^{h_2}
\f{t^{-2h}}{\Gamma(\beta-2h)}\dd h.
\end{equation}
Now we check the asymptotics of the integral above,
\begin{align}
&\f{\ln(t)}{t^{-2h_1}}\int_{h_1}^{h_2}\f{t^{-2H}}{\Gamma(\beta-2h)}\dd h
=\ln(t)\int_0^{h_2-h_1}\f{\e^{-2h\ln(t)}}{\Gamma(\beta-2h_1-2h)}\dd h
\nonumber\\
&=\int_0^{\ln(t)(h_2-h_1)}\f{\e^{-2h}}{\Gamma(\beta-2h_1-2h/\ln(t))}\dd h
\xrightarrow{t\to\infty}\f{1}{2\Gamma(\beta-2h_1)}.
\end{align}
Taking the limit $\epsilon\to 0$ and averaging over $Z$ proves point i).

For ii) let us first study the behaviour of the power law asymptotic itself,
\begin{align}
&\E\lt[\f{t^{-2(h_1+H')}}{Z\Gamma(\beta-2(h_1+H'))}\f{1}{t^{-2h_1}\ln(t)^{
-\alpha}}|Z\rt]
\nonumber\\
&=\f{1}{Zt^{-2h_1}\ln(t)^{-\alpha}}\int_0^{1-h_1} p_{H'}(h)\f{t^{-2(h_1+h)}}{
\Gamma(\beta-2(h_1+h))}\dd h
\nonumber\\
&=\f{1}{Z}\int_0^{\ln(t)(1-h_1)} p_H\lt(\f{h}{\ln t}\rt)\ln(t)^{\alpha-1}\f{
e^{-2h}}{\Gamma(\beta-2(h_1+h/\ln(t)))}\dd h
\nonumber\\
&\xrightarrow{t\to\infty}\f{1}{Z}\int_0^{\infty} h^{\alpha-1}\f{e^{-2h}}{
\Gamma(\beta-2h_1)}\dd h=\f{\Gamma(\alpha)}{Z2^\alpha\Gamma(\beta-2h_1)}.
\end{align}
Now, because of asymptotic \Ref{eq:MLasympt2} for every $\epsilon>0$ there exist
a $T_H$ such that for $t>T_H$
\begin{equation}\label{eq:MLineq}
\f{1-\epsilon}{Z\Gamma(\beta-2H)}t^{-2H}<E_{2H,\beta}(-Zt^{2H})<\f{1+\epsilon}{
Z\Gamma(\beta-2H)}t^{-2H}.
\end{equation}
This inequality holds for any $H$ in a closed interval $[h_1,h_2]$ and fixed
$Z$. As the Mittag-Leffler function and the power function are continuous with
respect to $H$ in this range, we can find $T$ sufficiently large such that this
inequality will hold for $t>T$ and all $H\in [h_1,h_2]$ simultaneously. Otherwise
we could take $T_k\to\infty$ and corresponding $H_k\in[h_1,h_2]$ for which
it does not hold and obtain a contradiction with continuity of $H\mapsto
E_{2H,\beta}(-Zt^{2H})$ or asymptotic \Ref{eq:MLasympt2} at an accumulation
point of the sequence $H_k$.

We may divide \Ref{eq:MLineq} by 
\begin{equation}
l(t)\equiv\f{\Gamma(\alpha)}{Z2^\alpha \Gamma(\beta-2h_1)}t^{-2h_1}\ln(t)^{-\alpha}
\end{equation}
and consider some large $t>T$ in order to obtain 
\begin{equation}
\label{eq:powerLawCovA}
1-\epsilon\le\liminf_{t\to\infty}\f{\E[E_{2H,\beta}(-Zt^{2H})|Z]}{l(t)},\quad
\limsup_{t\to\infty}\f{\E[E_{2H,\beta}(-Zt^{2H})|Z]}{l(t)}\le 1+\epsilon.
\end{equation}
Taking the limit $\epsilon\to0$ and averaging over $Z$ yields the desired asymptotic
\begin{equation}
\E[E_{2H,\beta}(-Zt^{2H})]\sim \E[l(t)]=\f{\E[Z^{-1}]\Gamma(\alpha)}{2^\alpha
\Gamma(\beta-2h_1)}t^{-2h_1}\ln(t)^{-\alpha}.
\end{equation}
\end{proof}

\end{document}